# High-Pressure Synthesis of Magnetic Neodymium Polyhydrides

*Di Zhou,[1] Dmitrii V. Semenok,[2] Hui Xie,[1] Xiaoli Huang,[1,\*] Defang Duan,[1] Alex Aperis,[3] Peter M. Oppeneer,[3] Michele Galasso,[2] Alexey I. Kartsev,[5,6] Alexander G. Kvashnin,[2,\*] Artem R. Oganov,[2,4,\*] and Tian Cui [7,1,\*]*

[1] State Key Laboratory of Superhard Materials, College of Physics, Jilin University, Changchun 130012, China

[2] Skolkovo Institute of Science and Technology, Skolkovo Innovation Center, 3 Nobel Street, Moscow 143026, Russia

[3] Department of Physics and Astronomy, Uppsala University, P.O. Box 516, SE-75120 Uppsala, Sweden

[4] International Center for Materials Discovery, Northwestern Polytechnical University, Xi'an 710072, China

[5] Computing Center of Far Eastern Branch of the Russian Academy of Sciences (CC FEB RAS), Khabarovsk, Russian Federation

[6] School of Mathematics and Physics, Queen's University Belfast, Belfast BT7 1NN, Northern Ireland, United Kingdom

[7] School of Physical Science and Technology, Ningbo University, Ningbo, 315211, China

**Corresponding Authors**

Dr. X. Huang, e-mail: huangxiaoli@jlu.edu.cn, Dr. A.G. Kvashnin, e-mail: A.Kvashnin@skoltech.ru, Prof. T. Cui, e-mail: cuitian@jlu.edu.cn, Prof. A.R. Oganov, email: a.oganov@skoltech.ru

**ABSTRACT**. The current search for room-temperature superconductivity is inspired by the unique properties of the electron-phonon interaction in metal superhydrides. Encouraged by the recently found highest-$T_C$ superconductor *fcc*-LaH$_{10}$, here we discover several superhydrides of another lanthanide — neodymium. We identify three novel metallic Nd-H phases at pressures range from 85 to 135 GPa: *I*4/*mmm*-NdH$_4$, *C*2/*c*-NdH$_7$, and *P*6$_3$/*mmc*-NdH$_9$, synthesized by laser-heating metal samples in NH$_3$BH$_3$ media for *in situ* generation of hydrogen. A lower trihydride $Fm\bar{3}m$-NdH$_3$ is found at pressures from 2 to 52 GPa. *I*4/*mmm*-NdH$_4$ and *C*2/*c*-NdH$_7$ are stable from 135 down to 85 GPa, and *P*6$_3$/*mmc*-NdH$_9$ from 110 to 130 GPa. Measurements of the electrical resistance of NdH$_9$ demonstrate a possible superconducting transition at ~ 4.5 K in *P*6$_3$/*mmc*-NdH$_9$. Our theoretical calculations predict that all the neodymium hydrides have antiferromagnetic order at pressures below 150 GPa and represent one of the first discovered examples of strongly correlated superhydrides with large exchange spin-splitting in the electron band structure (> 450 meV). The critical Néel temperatures for new neodymium hydrides are estimated using the mean-field approximation as about 4 K (NdH$_4$), 251 K (NdH$_7$) and 136 K (NdH$_9$).

**Introduction**

Since 1968, it is widely discussed that dense metallic hydrogen, if ever produced, could be a high-temperature superconductor [1]. The main reason is its very high Debye temperature (due to low atomic mass) and very strong electron-phonon coupling [2-3]. However, as creating metallic hydrogen requires immense pressures of ~500 GPa [4-6], a confirmation of high-$T_C$ superconductivity in pure hydrogen is still pending. Instead, in search for hydrogen-induced high-temperature superconductivity, most researchers have turned to hydrogen-rich hydrides [7]. Hydrides are promising to realize high-temperature superconductivity [8-10] under lower pressure compared to pure metallic hydrogen. Particularly, hydrides are convenient for experimental investigations because of a very high diffusion coefficient of atomic hydrogen, which facilitates the formation of new chemical compounds and makes it possible to perform laser-assisted chemical synthesis in diamond anvil cells in milliseconds [11].

Just a few years ago, a series of theoretical and experimental studies of various metal hydrides demonstrated that heavy elements relatively easily form metallic superhydrides (i.e. hydrides containing more hydrogen than expected based on atomic valences) at pressures below 200 GPa, such as *fcc*-LaH$_{10}$ [12-13], *P4/nmm*-LaD$_{11-12}$,[10] *P6$_3$/mmc*-ThH$_9$ and *Fm$\bar{3}$m*-ThH$_{10}$ [14], *P6$_3$/mmc*-UH$_7$, *Fm$\bar{3}$m*-UH$_8$ [15], *P6$_3$/mmc*-CeH$_9$ [16-17], and *P6$_3$/mmc*- and *F$\bar{4}$3m*-PrH$_9$ [18], whereas lighter atoms tend to form molecular semiconducting hydrides with low symmetry, for instance, LiH$_6$ [19] and NaH$_7$ [20]. Extremely high values of $T_C$ have been predicted for MgH$_6$ [21], CaH$_6$ [22] and YH$_{10}$ [23], but these compounds have not been synthesized yet.

One of the most important properties of high-symmetry polyhydrides is very strong electron-phonon interaction. The critical temperature (> 250 K [9-10]) and upper critical magnetic field ($H_C$ up to 140 T) achieved in *fcc*-LaH$_{10}$ have already surpassed by far the parameters of other known compounds, e.g. cuprates [24]. Analysis of theoretical results in the field shows that the increasing number of *d*- and *f*-electrons in the electron shells of a hydride-forming atom leads to the enhancement of magnetism, which can suppress superconductivity, and relative contribution of *d*- and *f*-orbitals to the total electron density of states at the Fermi level $N(E_F)$ associated with weakening of electron-phonon coupling [25]. Thus, the main driver of the study of lanthanide hydrides is their unusual crystal structures and possible interplay between the classical phonon-mediated superconductivity and magnetic ordering.

Currently, high-pressure experimental studies of lanthanide hydrides are not sufficiently developed, and further investigations of the synthetic pathways and crystal structure of these compounds will contribute to building a deeper understanding of the chemistry and physical properties of metal hydrides. Previously, Chesnut and Vohra [26] studied the crystal structure of metallic Nd at pressures up to 150 GPa and determined the phase sequence occurring as the pressure increases: *dhcp* → *fcc* → *dfcc* (*hR*24) → *hP*3 → monoclinic → α-U. Like other lanthanides, Nd can readily absorb hydrogen at high temperatures and form hydrides: $NdH_2$ with cubic close-packed and $NdH_3$ with hexagonal close-packed Nd sublattice were found at ambient pressure [27]. Continuing the investigations of lanthanide-hydrogen systems, we study here the crystal structures and properties of compounds in the Nd-H system in the pressure range of 0 to 140 GPa. Three novel superhydrides, $NdH_4$, $NdH_7$ and $NdH_9$ are synthesized, which display a fascinating combination of magnetism and superconductivity.

## Results and Discussion
### Stable phases predicted by theoretical calculations

Before the experiment, we carried out several independent variable-composition searches for stable compounds in the Nd-H system at 50, 100, and 150 GPa using the USPEX [28-30] code. The results of the structure search (Fig. 1) exhibit large differences depending on the inclusion or exclusion of spin–orbit coupling (SOC) and magnetism. Results with SOC and magnetism indicate that $Pm\bar{3}m$-NdH, tetrahydride $Immm$-$NdH_4$, cubic $Fm\bar{3}m$-$NdH_8$, and $F\bar{4}3m$-$NdH_9$ superhydrides are stable compounds at 150 GPa, while $C2/c$-$NdH_7$ lies ~0.05 eV/atom above the convex hull. $P6_3/mmc$-$NdH_9$, which is similar to recently discovered $CeH_9$ and $PrH_9$, lies 0.035 eV/atom above the convex hull (Fig. 1a). At 100 GPa (Fig. 1b, d) we see stabilization of the $I4/mmm$ modification of $NdH_4$ and appearance of $P6_3mc$-$NdH_8$ on the convex hull. At the same time, $C2/c$-$NdH_7$ becomes closer (~0.015 eV/atom) to the convex hull and becomes stable at about 80 GPa. The calculations at 50 GPa without SOC and magnetism show that $P6_3/mmc$-NdH, $P2_1/c$-$NdH_2$, $C2/m$-$NdH_3$, and $C2/c$-$NdH_7$ are stable (Fig. 1c). However, including of spin-orbit coupling leads to dramatic changes in the set of stable phases: $P2_1/c$-$NdH_2$ disappears, while $Fm\bar{3}m$-$NdH_3$ becomes thermodynamically stable. Molecular polyhydride $C2/c$-$NdH_7$ maintains its stability down to 50 GPa.

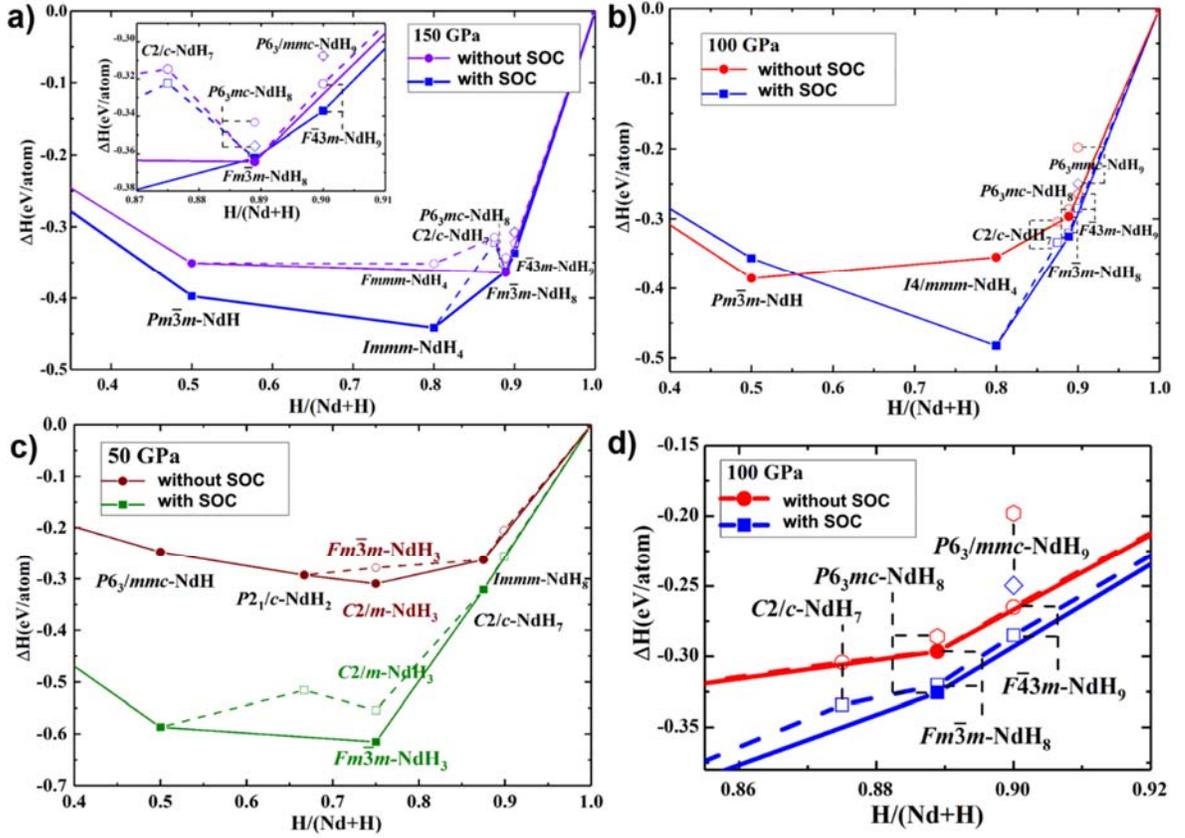

Fig. 1. Thermodynamic convex hulls of the Nd-H system calculated with and without spin-orbit coupling (SOC) and magnetism at (a) 150 GPa, (b) and (d) 100 GPa, and (c) 50 GPa. The enlarged part of Fig. 1b at 100 GPa is shown in (d).

**Experimental synthesis of atomic *I4/mmm*-NdH$_4$ and molecular *C2/c*-NdH$_7$**

Synthesis of neodymium hydrides was performed in a diamond anvil cell (DAC) that we denoted as Z1, containing a piece of Nd compressed in the NH$_3$BH$_3$ medium to 94 GPa and heated to 1700 K, according to the reaction: Nd + NH$_3$BH$_3$ → NdH$_x$ + *c*-BN [31-33]. The Raman signal of H$_2$ was detected at 4180.9 cm$^{-1}$, the corresponding pressure is 96 GPa. The rather complex diffraction pattern observed after laser heating shows that the reaction products are dominated by *C2/c*-NdH$_7$ with small impurity of tetragonal *I4/mmm*-NdH$_4$ (Fig. 2 and Fig. S1). Exploring this sample, we determined the experimental equation of state (EoS) of *C2/c*-NdH$_7$ and *I4/mmm*-NdH$_4$ in the pressure range from 85 to 135 GPa (Fig. 3b), and found close agreement with the EoS obtained by DFT calculations.

In $I4/mmm$-NdH$_4$, which is isostructural to the recently found $I4/mmm$-ThH$_4$ [14, 34], CeH$_4$ [17], and CaH$_4$ [35], the shortest H-H distance is $d_{min}$(H–H) = 1.55 Å at 100 GPa. Each Nd atom is bounded to 10 H atoms with $d$ (Nd–H) = 2.02-2.08 Å. The hydrogen sublattice in NdH$_4$ is represented by the atomic hydrogen with a high contribution of 1$s$-electrons to the electron density of states $N(E_F)$. The experimental cell parameters of discovered compounds are shown in Supplementary Table S4. Fitting the experimental pressure-volume data in the pressure range from 85 to 135 GPa by the 3$^{rd}$ order Birch-Murnaghan equation of state gives $V_{100}$ = 45.6(6) Å$^3$, $K_{100}$ = 500.(8) GPa, and $K_{100}'$ = 4.1(7).

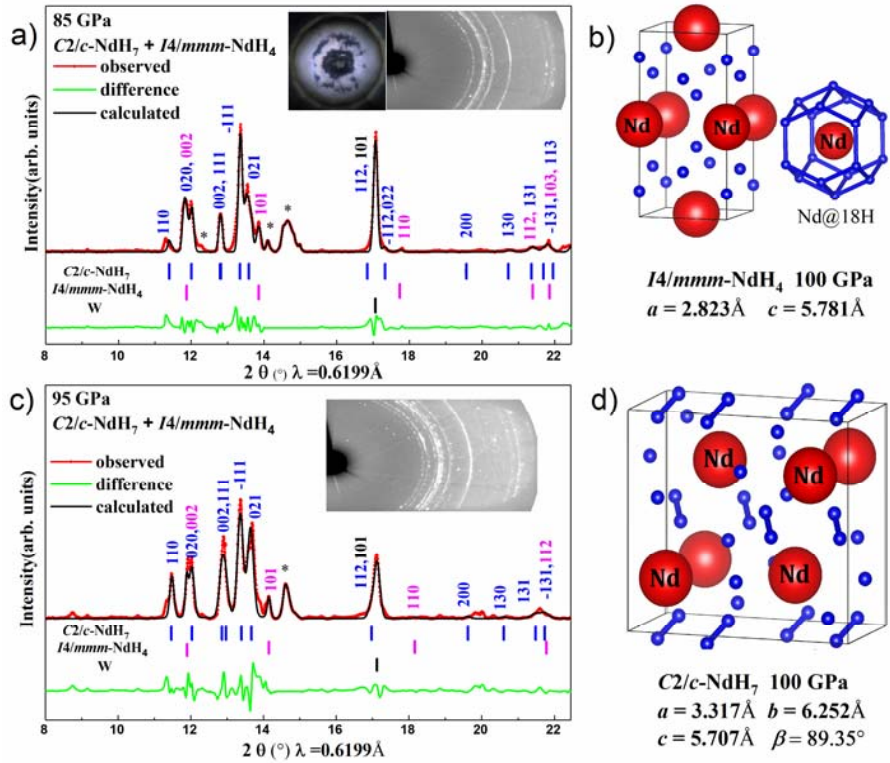

Fig. 2. Experimental X-ray diffraction (XRD) patterns and Le Bail refinement of $I4/mmm$-NdH$_4$ and $C2/c$-NdH$_7$ at (a) 85 GPa and (c) 95 GPa. The experimental data and model fit for the structure are shown in red and black, respectively; the residues are indicated in green, the unexplained peaks are marked by asterisks. The crystal structures of $I4/mmm$-NdH$_4$ and $C2/c$-NdH$_7$ at 100 GPa are shown in (b) and (d), respectively.

The experimentally discovered $C2/c$-NdH$_7$ structure, which is close to the recently predicted $C2/m$-AcH$_8$ [36] and $P2_1/m$-ThH$_7$ [34] at 100 GPa, contains quasimolecular hydrogen H$_2$ units with $d$(H–H) = 0.92Å. The remaining hydrogen (~30%) is in the atomic form. Each Nd atom is bounded to 17 H atoms with $d$(Nd-H) = 1.98-2.06 Å. The experimental cell parameters of this phase are shown in Supplementary Table S5. Fitting the experimental pressure-volume data in the pressure range from 85 to 135 GPa by the 3$^{rd}$ order Birch-Murnaghan equation of state with fixed $K_{100}' = 4$ gives $V_{100} = 118.1(3)$ Å$^3$, $K_{100} = 505.(3)$ GPa.

**Synthesis of superhydride $P6_3/mmc$-NdH$_9$**

Synthesis of higher neodymium hydrides was performed in Z2 diamond anvil cell with Nd particle compressed in the NH$_3$BH$_3$ medium to 113 GPa and heated to 1800 K. The Raman signal of H$_2$ was detected at 4142.9 cm$^{-1}$, the corresponding pressure is 115 GPa. The obtained diffraction pattern (Fig. 3a, Supplementary material Fig. 2b and Table S6) corresponds to hexagonal $P6_3/mmc$-NdH$_9$, the structure of which is very close to the previously described hexagonal $P6_3/mmc$-ThH$_9$ [14], UH$_9$ [15] and PrH$_9$ [18] (Fig. 3c). Both known theoretical investigations of the high-pressure chemistry of the Nd-H system [13, 25] claim the existence of cubic NdH$_8$ and NdH$_9$, while the experiment shows the presence of only one unexpected nonstoichiometric hexagonal $P6_3/mmc$-NdH$_9$ phase.

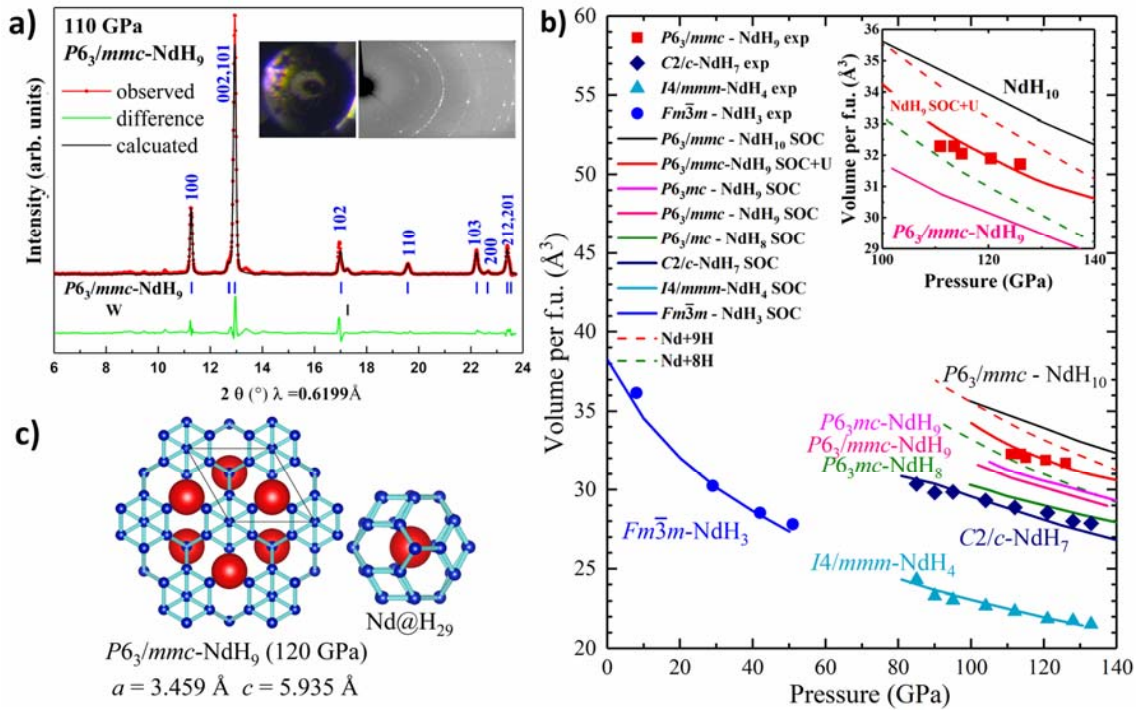

Fig. 3. (a) Experimental XRD pattern and Le Bail refinement of $P6_3/mmc$-NdH$_9$. The experimental data and model fit for the structure are shown in red and black, respectively; the residues are indicated in green. (b) The equation of state of the synthesized Nd-H phases; theoretical results include spin–orbit coupling and magnetism. Inset: The distinction between $P6_3mc$-NdH$_8$, $P6_3/mmc$- and $P6_3mc$-NdH$_9$, and $P6_3/mmc$-NdH$_{10}$ phases. (c) The crystal structure of $P6_3/mmc$-NdH$_9$ and parameters of the unit cell.

The calculated volumes of the ideal $P6_3/mmc$-NdH$_9$ structure are close to the experimental values, although the theoretical cell parameters have some deviations from the observed ones, for instance: $a$(exp) = 3.639 Å, $a$(theory) = 3.459 Å, $c$(exp) = 5.560 Å, $c$(theory) = 5.935 Å at 120 GPa. These deviations, as well as the stability of hexagonal NdH$_9$, can be almost completely explained by DFT+U calculations with $U$-$J$ = 5 eV, chosen by using linear-response calculations (inset in Fig. 3b, see Supplementary material for details) and introduced for describing correlation effects. Note that such values of $U$-$J$ are commonly used for modelling $f$-electrons in Nd [37-41]. Additional *ab initio* studies show that the $P6_3/mmc$-NdH$_9$ structure is dynamically stable (Supplementary material, Fig. S5).

Because of the visible difference in cell parameters, we also considered possible presence of additional hydrogen in the structure: NdH$_{9+x}$ (x=0-0.5), where the degree of nonstoichiometry (x) was determined by a linear interpolation of the dependence of the cell volume on the H content between 9 and 10 H atoms per Nd atom. In addition, we investigated $P6_3mc$-NdH$_8$ and $P6_3/mmc$-NdH$_{10}$ structures as possible candidates. NdH$_8$ is similar to predicted $P6_3mc$-PrH$_8$ and dynamically stable, lying on the convex hull at 100 GPa (Fig. 3b). However, its predicted volume $V_{120}$ (NdH$_8$) = 29.04 Å$^3$, is much smaller than the experimental value of 31.88 Å$^3$. The proposed $P6_3/mmc$-NdH$_{10}$ is thermodynamically and dynamically unstable, and its predicted cell volume differs significantly from the measured one (Fig. 3b).

NdH$_9$ has the same hexagonal structure as the whole family of reported hexagonal superhydrides: YH$_9$, CeH$_9$, PrH$_9$ and ThH$_9$. In $P6_3/mmc$-NdH$_9$, H$_{29}$ cages have the nearest H–H distance of 1.272 Å, the longest among all previously studied lanthanide polyhydrides at 120 GPa, while the nearest Nd–H distance at this pressure is 1.973 Å. The experimental cell parameters of the compound are listed in Supplementary Table S6. Fitting the experimental pressure-volume data in the range from 110 to 126 GPa using the 3$^{rd}$ Birch-Murnaghan equation of state with fixed $K_{100}'$ = 4 gives $V_{100}$ = 32.(9) Å$^3$, $K_{100}$ = 635.(9) GPa.

After the destruction of diamonds in the Z2 DAC, followed by decompression from 106 to 51 GPa, the recorded XRD patterns demonstrated presence of only one hydride phase: the metallic and magnetic $Fm\overline{3}m$-NdH$_3$ (Supplementary material, Fig. S2c) with the experimental cell parameter $a$ = 4.814 Å at 50 GPa, in agreement with the earlier predictions[13]. Fitting the experimental pressure-volume data for this phase in the pressure range from 7 to 50 GPa using the 3$^{rd}$ order Birch-Murnaghan equation of state gives $V_0$ =42.3 (3) Å$^3$, $K_0$ = 48.4 (9) GPa, and $K_0'$ = 4.5 (0).

**Measurements of the electrical resistance of $P6_3/mmc$-NdH$_9$**

The study of possible superconductivity in neodymium polyhydrides was performed in a Z3 cell with four 500 nm thick molybdenum electrodes prepared by magnetron sputtering (E = 200

Volts) and UV lithography on a diamond culet of 80 μm in diameter. The cell was loaded with a 30 μm particle of Nd and ammonium borane layer (thickness ~ 15-20 μm) and then compressed to a pressure of 110 GPa using *c*-BN/epoxy insulating gasket. After pulsed laser heating of the sample during 3 seconds (total, 4 runs) over 1600 K and subsequent cooling in a cryostat, the electrode system was partially damaged and several electrodes sequentially lost contact with the sample. In these cases, 3-electrode (pseudo van der Pauw) and 2-electrode schemes were used (Fig. 4). As a result of the resistance (R) measurements in the range of 1.6 - 230 K we have detected a sharp reproducible drop of R(T) at 4.5 ± 0.5 K, probably, caused by the superconducting transition in $NdH_9$. At the same time, no clear and reproducible transitions or resistance drops were detected above 5 K.

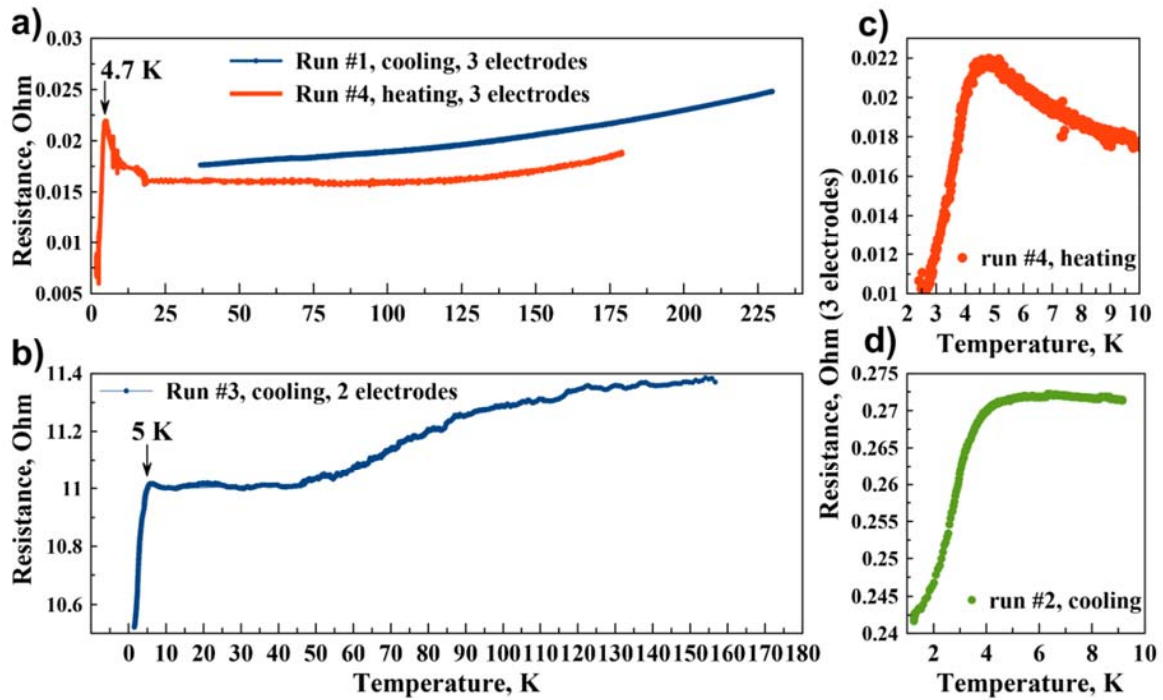

Fig. 4. Dependence of electrical resistance of the synthesized sample on the temperature: (a), (c) and (d) in 3-electrodes geometry; (b) in 2-electrodes geometry. Runs (#1-4) correspond to sequential steps of the laser heating.

To study the phase composition of the synthesized sample in electrical cell, X-ray diffraction was performed using Beijing Synchrotron Research Facilities (BSRF, China). The wavelength of the synchrotron radiation was λ = 0.6199 Å, exposure time - 10 minutes, the beam diameter ~ 50 µm. Due to presence of the electrodes, beam widening and its weak intensity, the X-ray diffraction pattern cannot be quantitatively interpreted, however, the obtained diffractogram (Fig. 5) qualitatively corresponds to a distorted hexagonal *P*6$_3$/*mmc*-NdH$_{9-x}$ mixed with tetragonal *I*4/*mmm*-NdH$_4$, which volume is close to the predicted one.

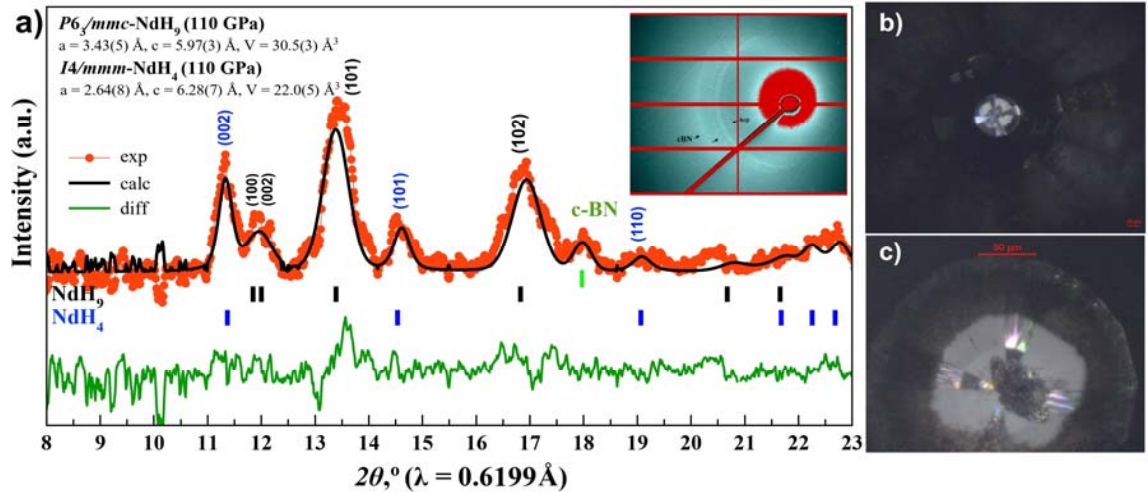

Fig. 5. (a) X-ray diffraction pattern and Le Bail refinement of the NdH$_9$ and NdH$_4$ structures found in the electrical cell at 110 GPa. Reflection from *c*-BN (a = 3.441 Å) corresponds to ~75 GPa; (b, c) optical microscopy of the culet of 4-electrodes electrical cell and sample after laser heating.

**Magnetic properties of neodymium hydrides**

*Ab initio* calculations show that *P*6$_3$/*mmc*-NdH$_9$ exhibits metallic properties and magnetism (Supplementary material, Fig. S6). The contribution of hydrogen atoms to $N(E_F)$ is very low: 7.44 eV$^{-1}$ f.u.$^{-1}$ (97%) comes from Nd, while hydrogen gives only 0.22 eV$^{-1}$f.u.$^{-1}$ (3%) at 120 GPa. It is expected that such high electron density at the Fermi level drives an instability against spontaneous magnetization (see Stoner criterion [42]). In spite of a clear analogy between Nd-H and Pr-H[18] systems (hexagonal XH$_9$ and tetragonal XH$_4$ polyhydrides), there is a significant difference in magnetic properties associated with the number of *f*-electrons. All Nd-H phases demonstrate strong magnetic properties (Fig. 6a and Fig. 6b), more pronounced than those of the corresponding praseodymium hydrides, except recently discovered *I*4/*mmm*-PrH$_4$[43] which also possesses AFM ordering at 100 GPa (see Supplementary Table S8, S9). This, probably, excludes the possibility of a classical *s*-wave superconductivity due to the large exchange spin-splitting in electron band structure (>450 meV). On the other hand, the coexistence of magnetism and relatively strong electron-phonon interaction leaves open the possibility for more complex mechanisms of superconductivity as found, e.g. in cuprates and Fe-containing pnictides [24, 44].

Simple spin-polarized calculations show that $NdH_7$ and $NdH_4$ maintain the magnetic moment of ~ 2.5-3.5 $\mu_B$ per Nd atom at low pressures, then start to slowly lose magnetism as pressure increases, while cubic $NdH_3$ keeps an almost constant magnetic moment, around 3 $\mu_B$ per Nd atom in the pressure range from 0 to 150 GPa. The magnetization of $NdH_9$ rapidly decreases with rising pressure and vanishes at ~230 GPa (Fig. 6a, b).

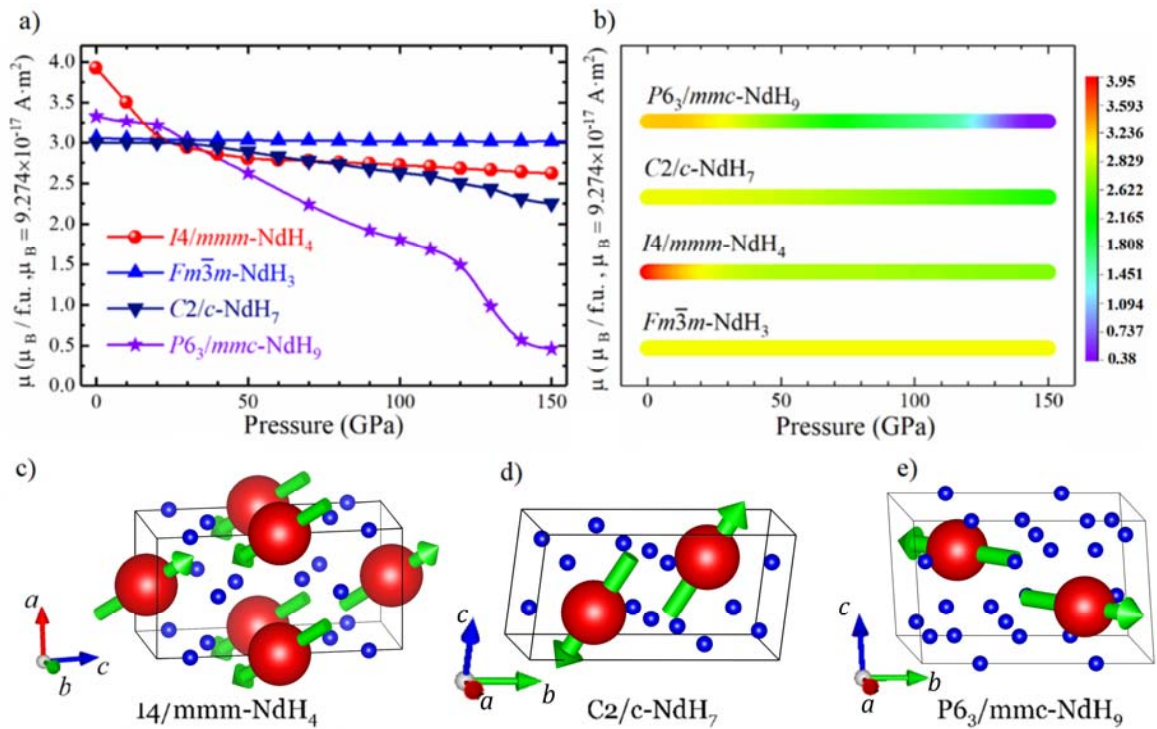

Fig. 6. Magnetic properties of Nd-H compounds. Magnetism of Nd hydrides at pressures up to 150 GPa. (a) Magnetic moments (per Nd atom) at various pressures and (b) magnetic map of the Nd-H system as a function of pressure. Computed spin configurations for 2 × 2 × 1 supercell of $NdH_4$ (c) and unit cells of $NdH_7$ (d) $NdH_9$ (e). Arrows indicate magnetic moments directions on the Nd-atoms.

To understand the exact spin configurations at finite pressure we have employed 10 different spin configurations using single-unit cells for NdH$_7$, NdH$_9$, and 13 spin configurations and 2 × 2 × 1 supercell - for NdH$_4$. Using non-collinear calculations, the magnetic anisotropy energy (MAE) for Nd-H phases along different directions has been computed. For calculations, each magnetic configuration has been calculated at fixed lattice parameters and corresponding pressure. The results summarized in Table 1 clearly indicate the antiferromagnetic (AFM) character of NdH$_4$, NdH$_7$ and NdH$_9$ compounds (Fig. 4 c-e, Supplementary Tables S8, S9).

We have identified that NdH$_4$, NdH$_7$ and NdH$_9$ compounds possess AFM collinear [112], [144] and [23$\bar{1}$] orders, respectively. The Néel temperature can be estimated using mean-field approximation $Z \cdot T_N^{MF} = \sum_{i,j} \frac{J_{i,j} S^2}{3k_B} \approx \min \frac{|E_{FM} - E_{AFM}|}{6k_B}$ as ~4 K (NdH$_4$), ~251 K (NdH$_7$) and ~136 K (NdH$_9$). As it will be demonstrated below, the Néel temperatures are significantly higher than expected superconducting $T_C$ for all neodymium hydrides.

**Electron-phonon interaction in neodymium hydrides**

The study of electron-phonon (*el-ph*) interaction and superconducting properties of neodymium hydrides is complicated because of two reasons: 1) possible interplay between magnetic ordering and *el-ph* interaction (see below) for which there is no relevant theory at this moment [24, 44], and 2) complex structure of the electronic density of states *N(E)* (or DOS) with multiple van Hove singularities (vHs) near the Fermi level (Figure S6, S11) caused by *f*-electrons of Nd. The latter makes us go beyond the commonly accepted "constant DOS approximation" [45] in calculations of the critical temperature, and take into account the exact structure of electron density of states.

One of the consequences of the described effect is a strong dependence of parameters of *el-ph* interaction on Gaussian DOS broadening (σ) parameter used in Quantum ESPRESSO (QE) (see Supplementary material) [46]. A similar situation was previously observed for praseodymium superhydrides $P6_3/mmc$-PrH$_9$ and $F\bar{4}3m$-PrH$_9$ [18], where results of QE calculations are strongly dependent on σ and the chosen Pr pseudopotential.

To solve this problem, we applied a method developed by Lie and Carbotte [47] and analyzed the solution of the linearized Eliashberg equations [48] taking into account the detailed structure of $N(E)$. As it was shown[47] the $N(E)$ affects the Eliashberg equations *via* the following integral (1):

$$\bar{N}(\overline{|\omega_l|}) = \frac{1}{\pi} \int_{-\infty}^{\infty} \frac{\overline{|\omega_l|}}{E^2 + \overline{|\omega_l|}^2} \frac{N(E)}{N(0)} dE \qquad (1)$$

the contribution of which is determined by the E ~0 (= $E_F$) region in $N(E)$. To illustrate this, we calculated this integral for the studied compounds: NdH$_4$, NdH$_7$ and NdH$_9$ (Figure S11). Our results show that the maximum positive influence of $N(E)$ on $T_C$ is for NdH$_7$ and NdH$_9$, while for NdH$_4$ this effect is negative.

Harmonic calculations using numerical solution of the isotropic set of Eliashberg equations and σ = 0.025 Ry (which gives good convergence) show that NdH$_9$ at 120 GPa has a quite small electron-phonon coupling constant λ = 0.43, $\omega_{log}$ = 602 K and formal $T_C$ = 4.3 K (μ*=0.1), close to experimental critical temperature. Contribution of the hydrogen sublattice to the DOS is very small (1-3 %) that is why the matrix elements of *el-ph* interaction are ~ 0. As it is shown in Fig. S11, accounting for the exact $N(E)$ structure leads to the increase of critical temperature. Calculations within the "constant DOS approximation" [45] led to a 10% lower $T_C$ = 3.7 K (μ*=0.1).

Combining the results of the VASP calculations with those of QE we have obtained a more reliable Eliashberg function (Fig. S13) to describe the electron-phonon interaction in NdH$_9$ (see Supplementary material). This $\alpha^2F(\omega)$ leads to much higher λ = 2.82, and lower $\omega_{log}$ = 272 K. The resulting critical temperature is 63 K (μ*=0.1) according to the Allen-Dynes formula[49]. However, consideration of antiferromagnetic ordering leads to an almost complete suppression of superconductivity (see below).

The electron-phonon interaction in the molecular *C2/c*-NdH$_7$ at 100 GPa is even weaker than in NdH$_9$: λ = 0.23, $\omega_{log}$ = 842 K, and formal $T_C$ (μ*=0.1) is ~ 0.01 K. This may be associated with the almost zero (~1%) contribution to $N(E_F)$ from hydrogen (Fig. S6(b)). Numerical solution of Eliashberg equations with accounting for the exact DOS structure $N(E)$ at formal $\mu^* = 0$ leads to maximum $T_C$ of 2.7 K while within the "constant DOS approximation" [45] we obtained much a lower value of 1.2 K.

Unexpectedly, a pronounced electron-phonon coupling is predicted for $I4/mmm$-NdH$_4$ (Fig. S10-11). At 100 GPa, this compound has $\lambda = 0.54$, $\omega_{log} = 857$ K that corresponds to $T_C = 13.3$ K ($\mu^*=0.1$). Due to the presence of a clear maximum of $N(E)$ close to $E_F$, the determination of superconducting $T_C$ strongly depends on the chosen σ-smoothing (Fig. S10-11). This has a negative effect on $T_C$ and within the "constant DOS approximation" [45] the critical temperature is 15.2 K.

**Eliashberg calculations with UppSC code**

In conventional superconductors, Cooper pairs are formed from electrons in time-reversed states and the relevant order parameter is proportional to the anomalous average $< c_{k\uparrow} c_{-k\downarrow} >$. Due to the opposite spin of the paired electrons, an applied magnetic field tends to break the Cooper pair apart and eventually at some critical field the superconducting state will be destroyed. In magnetic superconductors due to the underlying magnetic state, electrons feel such an effective magnetic field which has a detrimental influence on the superconducting state [50]. This effective magnetic field lifts the spin degeneracy, therefore inducing a spin-splitting "gap" in the electron energy dispersion. Electrons at the Fermi level now need to overcome this energy in order to form spin singlet Cooper pairs.

To consider the possible effect of the magnetic structure on electron-phonon interaction in $P6_3/mmc$-NdH$_9$ we analyzed the spin-resolved electronic band structure (Fig. S6) and computed solutions of the Eliashberg equations using the UppSC code (see Supplementary material). We found that the value of the spin-splitting at the Fermi level is too large (450 meV with $U-J =0$, and 890 meV at $U-J = 5$ eV) for any finite superconducting solution, as confirmed by our calculations which yield $T_C = 0$ K even for $\mu^* = 0$. Given that experiments find $T_C \sim 4.5$ K it is worthwhile to discuss possible scenarios. Aside from the potential presence of superconducting impurities (Mo,Nd)$C_xH_y$, it is possible that, within our current calculations we overestimated the spin-splitting ($h$) around the Fermi level. To investigate this scenario, we solved the Eliashberg equations for several values of $h$, $T$ and $\mu^*$. We found that for $\mu^* = 0.1$ and $h$ = 20 meV, $T_C \sim$ 5 K in agreement with experiment. Therefore, it may be that high-pressure $NdH_9$ could in principle be the first example material of the hydride family where superconductivity and AFM coexist, but at low temperature. Such a coexistence is not *a priori* to be excluded as has been shown e.g. in the case of unconventional AF superconductors [51]. Another possible scenario is that the low-temperature superconducting state may involve some form of a spin triplet superconducting order parameter. A possible candidate could be an equal spin triplet state [52]. Further investigations towards such an exciting direction would require solving the fully anisotropic Eliashberg equations[53], which goes beyond the scope of the present work.

## Conclusions

Following the remarkable discovery of $LaH_{10}$, here we synthesized three novel neodymium polyhydrides, *P*6$_3$/*mmc*-$NdH_9$, *C*2/*c*-$NdH_7$ and *I*4/*mmm*-$NdH_4$ through the *in situ* laser-assisted decomposition of $NH_3BH_3$ with the simultaneous absorption of the released hydrogen by metallic Nd. Hexagonal $NdH_9$ is the next member of the *P*6$_3$/*mmc* family of La, Ce, Pr, Th, and U nonahydrides. For all the synthesized phases, the equations of state and unit cell parameters are in satisfactory agreement with our DFT or DFT+U ($U$-$J$ = 5 eV) calculations. Preliminary measurements of the electrical resistance of $NdH_9$/$NdH_4$ sample point to a possible SC transition at 4.5±0.5 K at 110 GPa, and absence of superconductivity above 5 K. Although *P*6$_3$/*mmc*-$NdH_9$ has the highest H-content, the large spin-splitting in the electron band structure (> 450 meV) and antiferromagnetic ordering almost rule out classical *s*-wave superconductivity. Thus, the intensity of superconducting properties declines in the La-Ce-Pr-Nd series of superhydrides, while magnetic properties become more and more pronounced.

Theoretical calculations predict that all the neodymium hydrides exhibit strong magnetism at pressures below 150 GPa and possess collinear antiferromagnetic order, similar to *I4/mmm*-PrH$_4$, examined as a reference. The critical Néel temperatures for the newly synthesized neodymium hydrides were estimated using the mean-field approximation as about 4 K (NdH$_4$), 251 K (NdH$_7$), and 136 K (NdH$_9$).

## Acknowledgments

The authors thank the staff of the Shanghai and Beijing Synchrotron Radiation Facility. Authors express their gratitude to Bingbing Liu's group (Jilin University) for their help in the laser heating of samples. This work was supported by the National Key R&D Program of China (Grant No. 2018YFA0305900), the National Natural Science Foundation of China (Grant Nos. 11974133, 51720105007, 51632002, 11674122, 11574112, 11474127, and 11634004), the National Key Research and Development Program of China (Grant No. 2016YFB0201204), the Program for Changjiang Scholars and Innovative Research Team in University (Grant No. IRT_15R23), and the National Fund for Fostering Talents of Basic Science (Grant No. J1103202). A.R.O. thanks the Russian Science Foundation (Grant No. 19-72-30043). A.G.K., D.V.S. and A.I.K. thank the Russian Foundation for Basic Research (Grants No. 19-03-00100 and 18-29-11051 mk). A.A. and P.M.O. acknowledge support from the Swedish Research Council (VR), the Röntgen-Ångström cluster and from the Swedish National Infrastructure for Computing (SNIC). We thank Professor R. Akashi (University of Tokyo) for valuable discussions on the influence of precise density of states on critical temperature of superconductivity.

## Author Contributions:

X.H., A.G.K., and T.C. conceived this project. D.Z. and D.V.S. performed the experiment, D.V.S., H.X., D.D., A.R.O., A.G.K., M.G., A.I.K. and T.C. prepared the theoretical calculations and analysis. D.V.S., D.Z., X.H., A.R.O., and T.C. wrote and revised the paper. A.A. performed the calculations using UppSC code, A.A. and P.M.O. analyzed the results. All the authors discussed the results and offered useful inputs.

**Competing interests:**

The authors declare no competing interests.

**References**

1. Ashcroft, N. W., Metallic Hydrogen: A High-Temperature Superconductor? *Phys. Rev. Lett.* **1968,** *21* (26), 1748-1749.
2. McMahon, J. M.; Ceperley, D. M., High-temperature superconductivity in atomic metallic hydrogen. . *Phys. Rev. B* **2011,** *84* 144515.
3. Borinaga, M.; Errea, I.; Calandra, M.; Mauri, F.; Bergara, A., Anharmonic effects in atomic hydrogen: Superconductivity and lattice dynamical stability. *Phys. Rev. B* **2016,** *93* (17), 174308.
4. McMahon, J. M.; Morales, M. A.; Pierleoni, C.; Ceperley, D. M., The properties of hydrogen and helium under extreme conditions. *Rev. Mod. Phys.* **2012,** *84* (4), 1607-1653.
5. Azadi, S.; Monserrat, B.; Foulkes, W. M.; Needs, R. J., Dissociation of high-pressure solid molecular hydrogen: a quantum Monte Carlo and anharmonic vibrational study. *Phys. Rev. Lett.* **2014,** *112* (16), 165501.
6. McMinis, J.; III, R. C. C.; Lee, D.; Morales, M. A., Molecular to Atomic Phase Transition in Hydrogen under High Pressure. *Phys. Rev. Lett.* **2015,** *114*, 105305.
7. Ashcroft, N. W., Hydrogen dominant metallic alloys: high temperature superconductors? *Phys. Rev. Lett.* **2004,** *92* (18), 187002-1-187002-4.
8. Drozdov, A. P.; Eremets, M. I.; Troyan, I. A.; Ksenofontov, V.; Shylin, S. I., Conventional superconductivity at 203 kelvin at high pressures in the sulfur hydride system. *Nature* **2015,** *525* (7567), 73-6.
9. Somayazulu, M.; Ahart, M.; Mishra, A. K.; Geballe, Z. M.; Baldini, M.; Meng, Y.; Struzhkin, V. V.; Hemley, R. J., Evidence for Superconductivity above 260 K in Lanthanum Superhydride. *Phys. Rev. Lett.* **2019,** *122*, 027001.
10. Drozdov, A. P.; Kong, P. P.; Minkov, V. S.; Besedin, S. P.; Kuzovnikov, M. A.; Mozaffari, S.; Balicas, L.; Balakirev, F. F.; Graf, D. E.; Prakapenka, V. B.; Greenberg, E.; Knyazev, D. A.; Tkacz, M.; Eremets, M. I., Superconductivity at 250 K in lanthanum hydride under high pressures. *Nature* **2019,** *569* (7757), 528-531.
11. Goncharov, A. F.; Beck, P.; Struzhkin, V. V.; Hemley, R. J.; Crowhurst, J. C., Laser-heating diamond anvil cell studies of simple molecular systems at high pressures and temperatures. *Journal of Physics and Chemistry of Solids* **2008,** *69* (9), 2217-2222.

12. Geballe, Z. M.; Liu, H.; Mishra, A. K.; Ahart, M.; Somayazulu, M.; Meng, Y.; Baldini, M.; Hemley, R. J., Synthesis and Stability of Lanthanum Superhydrides. *Angew. Chem. Int. Ed.* **2017,** *129*, 6.
13. Peng, F.; Sun, Y.; Pickard, C. J.; Needs, R. J.; Wu, Q.; Ma, Y., Hydrogen Clathrate Structures in Rare Earth Hydrides at High Pressures: Possible Route to Room-Temperature Superconductivity. *Phys. Rev. Lett.* **2017,** *119* (10), 107001.
14. Semenok, D. V.; Kvashnin, A. G.; Ivanova, A. G.; Svitlyk, V.; Fominski, V. Y.; Sadakov, A. V.; Sobolevskiy, O. A.; Pudalov, V. M.; Troyan, I. A.; Oganov, A. R., Superconductivity at 161 K in thorium hydride ThH10: Synthesis and properties. *Materials Today* **2019**, 10.1016/j.mattod.2019.10.005.
15. I. A. Kruglov; A. G. Kvashnin; A. F. Goncharov; A. R. Oganov; S. S. Lobanov; N. Holtgrewe; S. Q. Jiang; V. B. Prakapenka; E. Greenberg; Yanilkin, A. V., Uranium polyhydrides at moderate pressuresPredict: Prediction, synthesis, and expected superconductivity. *Sci. Adv.* **2018,** *4*, eaat9776.
16. Salke, N. P.; Esfahani, M. M. D.; Zhang, Y.; Kruglov, I. A.; Zhou, J.; Wang, Y.; Greenberg, E.; Prakapenka, V. B.; Liu, J.; Oganov, A. R.; Lin, J.-F., Synthesis of clathrate cerium superhydride CeH9 at 80 GPa with anomalously short H-H distance. *Nat. Commun.* **2019,** *10*, 4453.
17. Li, X.; Huang, X.; Duan, D.; Pickard, C. J.; Zhou, D.; Xie, H.; Zhuang, Q.; Huang, Y.; Zhou, Q.; Liu, B.; Cui, T., Polyhydride CeH9 with an atomic-like hydrogen clathrate structure. *Nat. Commun.* **2019,** *10*, 3461.
18. Zhou, D.; Semenok, D. V.; Duan, D.; Xie, H.; Huang, X.; Chen, W.; Li, X.; Liu, B.; Oganov, A. R.; Cui, T., Superconducting Praseodymium Superhydrides. *Sci. Adv.* **2020**, DOI 10.1126/sciadv.aax6849.
19. Pépin, C.; Loubeyre, P.; Occelli, F.; Duma, P., Synthesis of lithium polyhydrides above 130 GPa at 300 K. *PNAS* **2015,** *112* (25), 7673-7676.
20. Struzhkin, V. V.; Kim, D. Y.; Stavrou, E.; Muramatsu, T.; Mao, H. K.; Pickard, C. J.; Needs, R. J.; Prakapenka, V. B.; Goncharov, A. F., Synthesis of sodium polyhydrides at high pressures. *Nat. Commun.* **2016,** *7*, 12267.
21. Feng, X.; Zhang, J.; Gao, G.; Liu, H.; Wang, H., Compressed Sodalite-like MgH6 as a Potential High-temperature Superconductor. *RSC Adv.* **2015,** *5* (73), 59292-59296.
22. Wang, H.; Tse, J. S.; Tanaka, K.; Iitaka, T.; Ma, Y., Superconductive sodalite-like clathrate calcium hydride at high pressures. *PNAS* **2012,** *109* (17), 6463-6466.
23. Liu, H.; Naumov, I. I.; Hoffmann, R.; Ashcroft, N. W.; Hemley, R. J., Potential high-Tc superconducting lanthanum and yttrium hydrides at high pressure. *PNAS* **2017,** *114* (27), 5.
24. Keimer, B.; Kivelson, S. A.; Norman, M. R.; Uchida, S.; Zaanen, J., From quantum matter to high-temperature superconductivity in copper oxides. *Nature* **2015,** *518* (7538), 179-86.
25. Semenok, D. V.; Kruglov, I. A.; Kvashnin, A. G.; Oganov, A. R., On Distribution of Superconductivity in Metal Hydrides. *arXiv:1806.00865* **2018**.
26. Chesnut, G. N.; Vohra, Y. K., a-uranium phase in compressed neodymium metal *Phys. Rev. B.* **2000,** *61*, R3768-R3771.

27. Vajda, P.; Daou, J. N., Rare Earths-Hydrogen. *Solid State Phenomena* **1996,** *49-50*, 71-158.
28. Oganov, A. R.; Glass, C. W., Crystal structure prediction using ab initio evolutionary techniques: Principles and applications. *J. Chem. Phys.* **2006,** *124*, 244704.
29. Oganov, A. R.; Lyakhov, R. O.; Valle, M., How Evolutionary Crystal Structure Prediction Works-and Why. *Acc. Chem. Res.* **2011,** *44*, 227-237.
30. Lyakhov, A. O.; Oganov, A. R.; Stokes, H. T.; Zhu, Q., New developments in evolutionary structure prediction algorithm USPEX. *Computer Phy. Commun.* **2013,** *184* (4), 1172-1182.
31. Chellappa, R. S.; Somayazulu, M.; Struzhkin, V. V.; Autrey, T.; Hemley, R. J., Pressure-induced complexation of NH3BH3–H2. *J. Chem. Phys.* **2009,** *131*, 224515.
32. Song, Y., New perspectives on potential hydrogen storage materials using high pressure. *Phys. Chem. Chem. Phys.* **2013,** *15* (35), 14524-47.
33. Potter, R. G.; Somayazulu, M.; Cody, G.; Hemley, R. J., High Pressure Equilibria of Dimethylamine Borane, Dihydridobis(dimethylamine)boron(III) Tetrahydridoborate(III), and Hydrogen. *The Journal of Physical Chemistry C* **2014,** *118* (14), 7280-7287.
34. Kvashnin, A. G.; Semenok, D. V.; Kruglov, I. A.; Wrona, I. A.; Oganov, A. R., High-Temperature Superconductivity in Th-H System at Pressure Conditions. *ACS Appl. Mater. Interfaces* **2018,** *10*, 43809-43816.
35. Wu, G.; Huang, X.; Xie, H.; Li, X.; Liu, M.; Liang, Y.; Huang, Y.; Duan, D.; Li, F.; Liu, B.; Cui, T., Unexpected calcium polyhydride CaH4: A possible route to dissociation of hydrogen molecules. *J. Chem. Phys.* **2019,** *150*, 044507.
36. Semenok, D. V.; Kvashnin, A. G.; Kruglov, I. A.; Oganov, A. R., Actinium Hydrides AcH10, AcH12, and AcH16 as High-Temperature Conventional Superconductors. *J. Phys. Chem. Lett.* **2018,** *9* (8), 1920-1926.
37. Zaari, H.; Boujnah, M.; El hachimi, A. G.; Benyoussef, A.; El Kenz, A., Electronic structure and X-ray magnetic circular dichroic of Neodymium doped ZnTe using the GGA + U approximation. *Computational Materials Science* **2014,** *93*, 91-96.
38. Reshak, A. H.; Piasecki, M.; Auluck, S.; Kityk, I. V.; Khenata, R.; Andriyevsky, B.; Cobet, C.; Esser, N.; Majchrowski, A.; Swirkowicz, M.; Diduszko, R.; Szyrski, W., Effect of U on the Electronic Properties of Neodymium Gallate (NdGaO3): Theoretical and Experimental Studies. *J. Phys. Chem. B* **2009,** *113*, 15237–15242.
39. Shankar, A.; Rai, D. P.; Thapa, R. K., Structural, electronic, magnetic and optical properties of neodymium chalcogenides using LSDA+Umethod. *Journal of Semiconductors* **2012,** *33* (8), 082001.
40. Kozub, A. L.; Shick, A. B.; Máca, F.; Kolorenč, J.; Lichtenstein, A. I., Electronic structure and magnetism of samarium and neodymium adatoms on free-standing graphene. *Phys. Rev. B* **2016,** *94* (12).
41. Morice, C.; Artacho, E.; Dutton, S. E.; Kim, H. J.; Saxena, S. S., Electronic and magnetic properties of superconducting LnO1-x F x BiS2 (Ln = La, Ce, Pr, and Nd) from first principles. *Journal of physics. Condensed matter : an Institute of Physics journal* **2016,** *28* (34), 345504.
42. Stoner, E. G., Collective electron ferromagnetism. *Proc. Roy. Soc.* **1938,** *A165*, 372.
43. Peña-Alvarez, M.; Binns, J.; Hermann, A.; Kelsall, L. C.; Dalladay-Simpson, P.; Gregoryanz, E.; Howie, R. T., Praseodymium polyhydrides synthesized at high temperatures and pressures. *Phys. Rev. B* **2019,** *100*, 184109.
44. Johnston, D. C., The puzzle of high temperature superconductivity in layered iron pnictides and chalcogenides. *Advances in Physics* **2010,** *59* (6), 803-1061.
45. Sano, W.; Koretsune, T.; Tadano, T.; Akashi, R.; Arita, R., Effect of Van Hove singularities on high-T_superconductivity in H3S. *Phys. Rev. B* **2019,** *93*, 094525.
46. Giannozzi, P.; Baroni, S.; Bonini, N.; Calandra, M.; Car, R.; Cavazzoni, C.; Ceresoli, D.; Chiarotti, G. L.; Cococcioni, M.; Dabo, I.; Corso, A. D.; Gironcoli, S. d.; Fabris, S.; Fratesi, G.; Gebauer, R.; Gerstmann, U.; Gougoussis, C.; Kokalj, A.; Lazzeri, M.; Martin-Samos, L.; Marzari, N.; Mauri, F.; Mazzarello, R.; Paolini, S.; Pasquarello, A.; Paulatto, L.; Sbraccia, C.; Scandolo, S.; Sclauzero, G.; Seitsonen, A. P.; Smogunov, A.; Umari, P.; Wentzcovitch, R. M., QUANTUM ESPRESSO: a modular and open-source software project for quantum simulations of materials. *J. Phys.: Condens. Matter* **2009,** *21* (39), 395502.
47. Lie, S. G.; Carbotte, J. P., Dependence of Tc on electronic density of states. *Sol. State. Comm.* **1978,** *26* (8), 511-514.
48. Eliashberg, G. M., Interactions between Electrons and Lattice Vibrations in a Superconductor. *JETP* **1959,** *11* (3), 696-702.
49. Allen, P. B.; Dynes, R. C., Transition temperature of strong-coupled superconductors reanalyzed. *Phys. Rev. B* **1975,** *12* (3), 905-922.
50. Fulde, P.; Zwicknagl, G., Antiferromagnetic superconductors (invited). *J. Appl. Phys.* **1982,** *53* (11), 8064-8069.
51. Aperis, A.; Varelogiannis, G.; Littlewood, P. B.; Simons, B. D., Coexistence of spin density wave, d-wave singlet and staggered π-triplet superconductivity - IOPscience. *Journal of Physics: Condensed Matter* **2008,** *20* (43), 434235.
52. Sigrist, M.; Ueda, K., Phenomenological theory of unconventional superconductivity. *Rev. Mod. Phys.* **1991,** *63*, 239.
53. Aperis, A.; Maldonado, P.; Oppeneer, P. M., Ab initio theory of magnetic-field-induced odd-frequency two-band superconductivity in MgB2. *Phys. Rev. B* **2015,** *92*, 054516.

# Supplementary Material

# for

# High-Pressure Synthesis of Magnetic Neodymium Polyhydrides

*Di Zhou,[1] Dmitrii V. Semenok,[2] Hui Xie,[1] Xiaoli Huang,[1,*] Defang Duan,[1] Alex Aperis,[3] Peter M. Oppeneer,[3] Michele Galasso,[2] Alexey I. Kartsev,[5,6] Alexander G. Kvashnin,[2,*] Artem R. Oganov,[2,4,*] and Tian Cui[7,1*]*

[1] State Key Laboratory of Superhard Materials, College of Physics, Jilin University, Changchun 130012, China

[2] Skolkovo Institute of Science and Technology, Skolkovo Innovation Center, 3 Nobel Street, Moscow 143026, Russia

[3] Department of Physics and Astronomy, Uppsala University, P.O. Box 516, SE-75120 Uppsala, Sweden

[4] International Center for Materials Discovery, Northwestern Polytechnical University, Xi'an 710072, China

[5] Computing Center of Far Eastern Branch of the Russian Academy of Sciences (CC FEB RAS), Khabarovsk, Russian Federation

[6] School of Mathematics and Physics, Queen's University Belfast, Belfast BT7 1NN, Northern Ireland, United Kingdom

[7] School of Physical Science and Technology, Ningbo University, Ningbo, 315211, China

*Corresponding authors: huangxiaoli@jlu.edu.cn, A.Kvashnin@skoltech.ru, cuitian@jlu.edu.cn and a.oganov@skoltech.ru

# CONTENTS

## Experimental methods

The neodymium powder with 99.99% purity was purchased from Alfa Aesar. We performed laser heating of two diamond anvil cells (100-μm culets) loaded with Nd and ammonia borane ($NH_3BH_3$) purified by sublimation. The thickness of the tungsten gasket was 20±2 μm. Heating was carried out by pulses of an infrared laser (1 μm, Nd: YAG). Temperature measurements were carried out by detection of the black-body radiation using the Planck formula on Mar345 detector. The pressure was determined by the Raman signal of diamond [1] using Horiba LabRAM HR800 Ev spectrometer with the exposure time of 60 seconds. The X-ray diffraction patterns of all samples studied in diamond anvil cells were recorded on BL15U1 synchrotron beamline [2] at Shanghai Synchrotron Research Facility (SSRF, China) using a focused (5×12 μm) monochromatic X-ray beam (20 keV, 0.6199 Å) and a two-dimensional Mar165 CCD detector. The experimental X-ray diffraction images were analyzed and integrated using Dioptas software package (version 0.4) [3]. The full profile analysis of the diffraction patterns and calculation of the unit cell parameters were performed in Materials Studio [4] and JANA2006 [5] using the Le Bail method [6].

## Computational details

We calculated the equations of state (EoS) for $NdH_3$, $NdH_4$, $NdH_7$, and $NdH_9$ phases and compared them with the experimental ones. In this work, we did not use the DFT+U correction. However, performing test DFT+U calculations by employing $U-J = 5$ eV we noticed its significant effect on the unit-cell volumes by about 3 % increase. To calculate the equations of state, we performed structure relaxations of phases at various pressures using density functional theory (DFT) [7-8] within the generalized gradient approximation (Perdew-Burke-Ernzerhof functional) [9] and the projector-augmented wave method [10-11] as implemented in the VASP code [12-14]. The plane wave kinetic energy cutoff was set to 1000 eV, and the Brillouin zone was sampled using Γ-centered $k$-points meshes with the resolution of 2π×0.05 Å$^{-1}$. Projected electron density of states was calculated by the l,m decomposed charge obtained *via* the integration of the charge density over the volumes calculated using Wigner-Seitz radius for each atom type. Non-collinear magnetism calculations have been carried out by including the spin-orbit coupling as implemented in VASP by Hobbs et. al [15].

The obtained dependences of the unit cell volume on pressure were formally fitted by the 3$^{rd}$ Birch-Murnaghan equation [16] to determine, using the EOSfit7 software [17], the main parameters of the EoS: the equilibrium volume $V_0$, bulk modulus $K_0$, and derivative of bulk modulus with respect to pressure $K'$. We also calculated phonon densities of states for the studied materials using the finite displacement method (VASP [18] and PHONOPY [19]).

Calculations of phonons, electron-phonon coupling, and superconducting $T_C$ were carried out with QUANTUM ESPRESSO (QE) package [20] using density-functional perturbation theory [21], employing the plane-wave pseudopotential method and Perdew-Burke-Ernzerhof exchange-correlation functional [9]. In our ab initio calculations of the electron-phonon coupling (EPC) parameter $λ$, the first Brillouin zone was sampled using 2×2×2 and 3×3×3 q-points meshes and a denser 16×16×16 $k$-points mesh (with the smearing σ = 0.025 Ry that approximates the zero-width limits in the calculation of λ). The critical temperature $T_C$ was calculated using the Allen-Dynes (A-D) formula [22] because for a low-coupling limit the results of the exact solution of the Eliashberg equation have almost no difference from the results of the much simpler A-D or McMillan formulas.

Verification of our choice of $U-J$ value at high pressure was done by using linear-response calculations with 3×3×3 q-points mesh in the frame of the plane-waves method based on the pseudopotential approach implemented in the Quantum ESPRESSO code [20]. In order to verify the portability and reproducibility of our results three different types of pseudopotential have been employed, namely ultrasoft potential based on the GGA-PBESOL, PAW potentials based on the GGA-PBESOL and PAW GGA-PBE approximations. The obtained values of the effective $U-J$ on-site interaction are consistent for all potentials and are the following: 4.98 eV ($NdH_9$ at 120 GPa), 4.18 eV ($NdH_7$ at 100 GPa) and 5.04 eV ($NdH_4$ at 100 GPa).

# Structural information

**Table S1.** The crystal structure of the discovered Nd-H phases.

| Phase | Pressure, GPa | Lattice parameters | Coordinates | | | |
|---|---|---|---|---|---|---|
| $P6_3/mmc$-NdH$_9$ | 120 | $a = b = 3.45888$ Å<br>$c = 5.93500$ Å<br>$\alpha = \beta = 90°$<br>$\gamma = 120°$ | Nd (2d) | 0.33333 | 0.66667 | 0.75000 |
| | | | H (12k) | 0.17856 | 0.35711 | 0.44156 |
| | | | H (4e) | 0.00000 | 0.00000 | -0.16525 |
| | | | H (2c) | 0.33333 | 0.66667 | 0.25000 |
| $C2/c$-NdH$_7$ | 100 | $a = 3.31774$ Å<br>$b = 6.25226$ Å<br>$c = 5.70716$ Å<br>$\alpha = \gamma = 90°$<br>$\beta = 89.354°$ | Nd (4e) | 0.00000 | 0.17130 | 0.25000 |
| | | | H (8f) | -0.37938 | 0.38789 | 0.40131 |
| | | | H (8f) | -0.47584 | 0.14567 | 0.03948 |
| | | | H (8f) | -0.27447 | 0.08499 | -0.05173 |
| | | | H (4e) | 0.00000 | 0.48949 | 0.25000 |
| $I4/mmm$-NdH$_4$ | 100 | $a = b = 2.82340$ Å<br>$c = 5.78080$ Å<br>$\alpha = \beta = \gamma = 90°$ | Nd (2b) | 0.00000 | 0.00000 | -0.50000 |
| | | | H (4e) | 0.00000 | 0.00000 | -0.13964 |
| | | | H (4d) | 0.00000 | -0.50000 | -0.25000 |
| $Fm\bar{3}m$-NdH$_3$ | 50 | $a = 4.78034$ Å<br>$\alpha = \beta = \gamma = 90°$ | Nd (4b) | -0.50000 | -0.50000 | 0.50000 |
| | | | H (4a) | 0.00000 | 0.00000 | 0.00000 |
| | | | H (8c) | -0.25000 | -0.25000 | 0.25000 |

**Table S2.** The experimental parameters of DACs.

| #cell | Pressure of synthesis, GPa | Gasket | Sample size, μm | Composition/load |
|---|---|---|---|---|
| Z1 | 90 | W | 18 | Nd/BH$_3$NH$_3$ |
| Z2 | 113 | W | 14 | Nd/BH$_3$NH$_3$ |
| Z3 | 110 | $c$-BN/epoxy | 30 | Nd/BH$_3$NH$_3$ |

**Table S3.** The experimental EoS parameters of pure elemental Nd at pressures of 0 to 150 GPa fitted by the modified universal equation of state (MUEOS) [23].

$$\ln P + \ln \frac{(V/V_0)^{2/3}}{3(1-(V/V_0)^{1/3})} = \ln B_0 + 1.5(B'_0 - 1) \cdot (1 - (V/V_0)^{1/3}) + \beta(1 - (V/V_0)^{2/3}) \quad (S1)$$

| | $dhcp$-Nd | $C2/m$-Nd | $Cmcm$-Nd |
|---|---|---|---|
| $V_0$ (GPa) | 34.165 | 34.165 | 34.165 |
| $K_0$ (GPa) | 25.38 | 24.05 | 6.141 |
| $K'$ | 3.119 | 3.225 | 7.351 |
| $\beta$ | 4.075 | 3.975 | 0 |

**Table S4.** The experimental cell parameters of $I4/mmm$-NdH$_4$ (Z=2).

| Pressure, GPa | Volume, Å$^3$ | a=b, Å | c, Å | V$_{DFT}$, Å$^3$ |
|---|---|---|---|---|
| 85 | 48.48(3) | 2.843(9) | 5.994(1) | 48.08 |
| 90 | 46.62(2) | 2.790(4) | 5.986(9) | 47.35 |
| 95 | 46.08(1) | 2.774(0) | 5.987(8) | 46.76 |
| 104 | 45.37(3) | 2.762(8) | 5.943(3) | 45.70 |
| 112 | 44.69(4) | 2.742(9) | 5.939(8) | 44.81 |
| 121 | 43.73(1) | 2.720(1) | 5.910(6) | 43.84 |
| 128 | 43.51(1) | 2.712(2) | 5.915(2) | 43.16 |
| 133 | 43.11(0) | 2.707(0) | 5.878(0) | 42.76 |

**Table S5.** The experimental cell parameters of $C2/c$-NdH$_7$ (Z=4).

| Pressure, GPa | Volume, Å$^3$ | a, Å | b, Å | c, Å | β,° | V$_{DFT}$, Å$^3$ |
|---|---|---|---|---|---|---|
| 85 | 121.5(2) | 3.667(5) | 5.942(1) | 5.574(2) | 87.23(3) | 122.66 |
| 90 | 119.3(3) | 3.636(7) | 5.939(2) | 5.524(4) | 88.34(7) | 121.39 |
| 95 | 119.4(5) | 3.642(5) | 5.918(6) | 5.545(3) | 87.56(9) | 119.98 |
| 104 | 117.3(1) | 3.630(4) | 5.895(9) | 5.484(8) | 87.82(0) | 117.10 |
| 113 | 115.5(2) | 3.617(5) | 5.847(5) | 5.463(2) | 87.82(3) | 114.98 |
| 121 | 114.2(2) | 3.601(9) | 5.806(6) | 5.462(6) | 87.75(4) | 112.41 |
| 127 | 112.0(1) | 3.569(0) | 5.795(5) | 5.419(3) | 88.30(2) | 110.48 |
| 133 | 111.5(0) | 3.559(6) | 5.791(9) | 5.412(2) | 87.90(0) | 109.21 |

**Table S6.** The experimental cell parameters of $P6_3/mmc$-NdH$_9$ and $Fm\overline{3}m$-NdH$_3$ (Z=2).

| $P6_3/mmc$-NdH$_9$ | | | | | | $Fm\overline{3}m$-NdH$_3$ | | | |
|---|---|---|---|---|---|---|---|---|---|
| Pressure, GPa | a, Å | c, Å | V per f.u., Å$^3$ | V$_{DFT}$ per f.u. Å$^3$ | V$_{DFT+U}$ per f.u. Å$^3$ | Pressure, GPa | a, Å | V, Å$^3$ | V$_{DFT}$, Å$^3$ |
| 111 | 3.645(3) | 5.610(0) | 32.3(1) | 30.76 | 32.94 | 8 | 5.249(4) | 36.16(3) | 35.75 |
| 113 | 3.650(2) | 5.595(1) | 32.3(1) | 30.59 | 32.76 | 29 | 4.946(1) | 30.25(1) | 31.43 |
| 115 | 3.641(8) | 5.580(4) | 32.0(1) | 30.47 | 32.58 | 42 | 4.851(1) | 28.54(1) | 29.65 |
| 120 | 3.639(4) | 5.560(3) | 31.9(2) | 30.12 | 32.14 | 51 | 4.809(2) | 27.81(1) | 28.58 |
| 126 | 3.629(4) | 5.558(2) | 31.7(2) | 29.75 | 31.64 | | | | |

**Table S7.** The calculated EoS parameters in the 3$^{rd}$ order Birch–Murnaghan equation for all studied Nd-H phases. For NdH$_4$ (Z=2), NdH$_7$ (Z=4) and NdH$_9$ $V_0$, $K_0$, $K'$ correspond to 100 GPa.

| | $Fm\overline{3}m$-NdH$_3$ | $I4/mmm$-NdH$_4$ | $C2/c$-NdH$_7$ | $P6_3/mmc$-NdH$_9$ |
|---|---|---|---|---|
| $V_0$ (GPa) | 42.3(3) | 45.7(2) | 118.3(1) | 32.8(1) |
| $K_0$ (GPa) | 48.4(9) | 525 (±15) | 522 (±23) | 719 (±41) |
| $K'$ | 4.5(0) | 2.7 (±1.2) | 2.4 (±1.5) | 4* |

*fixed K'=4 was used due to the insufficient pressure range

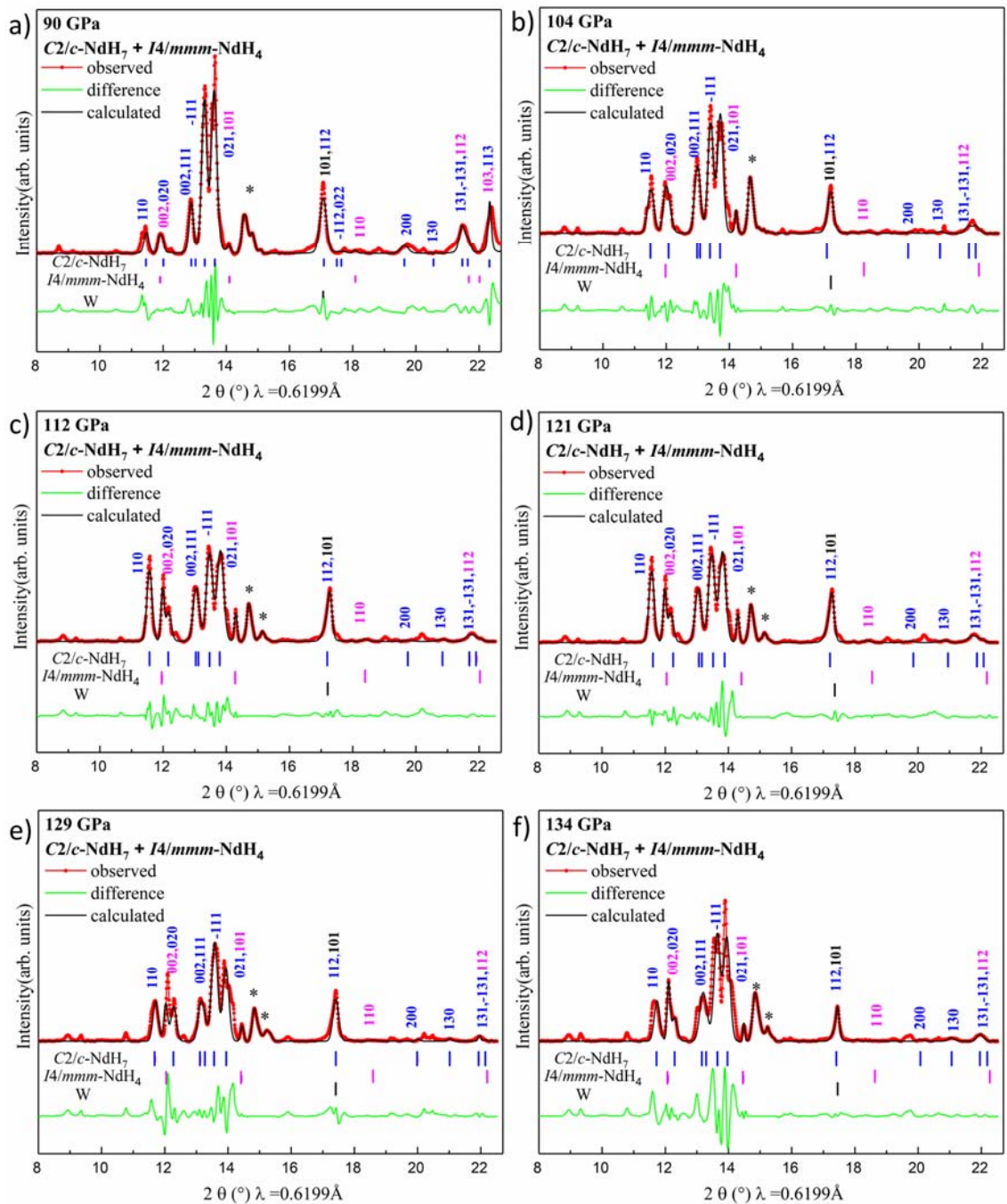

**Fig. S1.** Experimental XRD patterns and Le Bail refinements of *I*4/*mmm*-NdH$_4$ and *C*2/*c*-NdH$_7$ synthesized in Z1 cell at **(a)** 90 GPa, **(b)** 104 GPa, **(c)** 112 GPa, **(d)** 121 GPa, **(e)** 129 GPa, and **(f)** 134 GPa. The experimental data and model fit for the structure are shown in red and black, respectively; the residues are indicated in green.

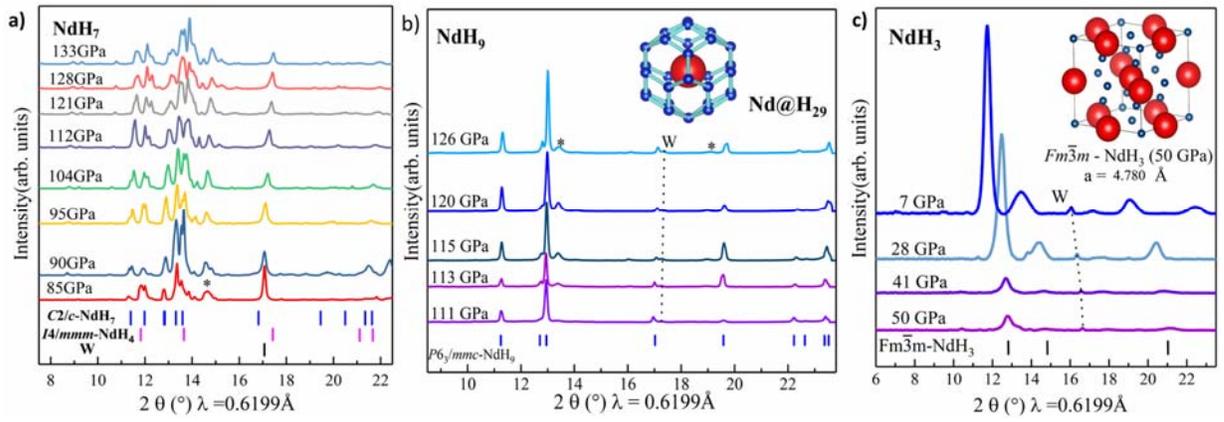

**Fig. S2. (a)** Experimental XRD patterns and Le Bail refinements of $I4/mmm$-NdH$_4$ and $C2/c$-NdH$_7$ phases synthesized in Z1 cell. **(b)** The XRD patterns of NdH$_9$ in the pressure range of 110 to 130 GPa and H-cage (H$_{29}$) structure of $P6_3/mmc$-NdH$_9$ at 120 GPa. Unclear weak reflections are marked by asterisks. **(c)** The experimental XRD patterns obtained from Z2 cell during the reduction of pressure from ~ 50 to 7 GPa.

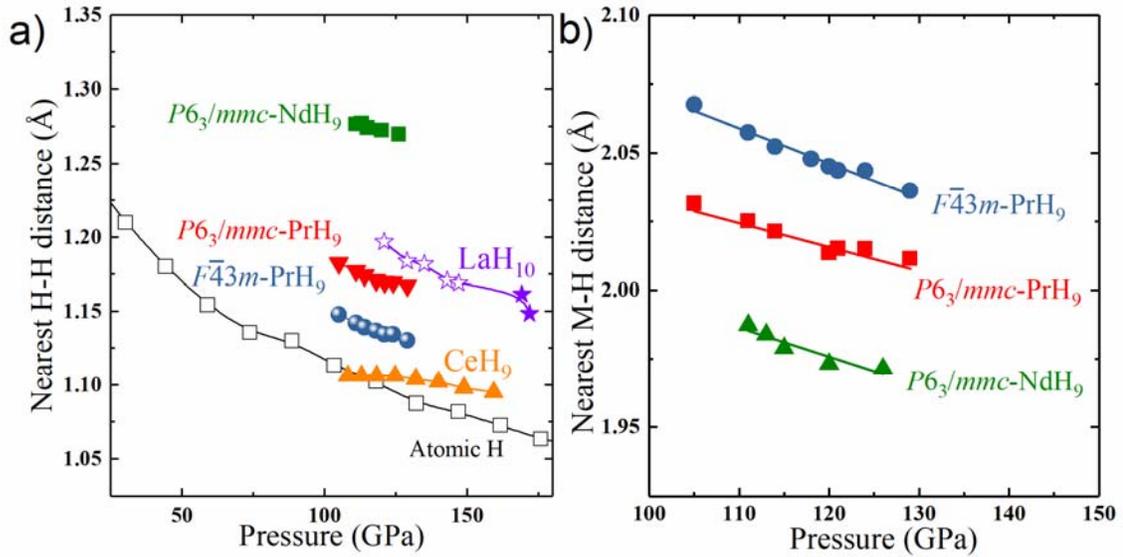

**Fig. S3. (a)** Pressure dependence of the nearest H–H distances for NdH$_9$, two phases of PrH$_9$ (Ref. 17, main text), LaH$_{10}$ (Ref. 12, main text), CeH$_9$ (Ref. 16, main text), and atomic H [24]. **(b)** Nearest Nd–H and Pr–H distances as a function of pressure calculated from the experimental cell parameters.

# Electron and phonon properties of neodymium hydrides

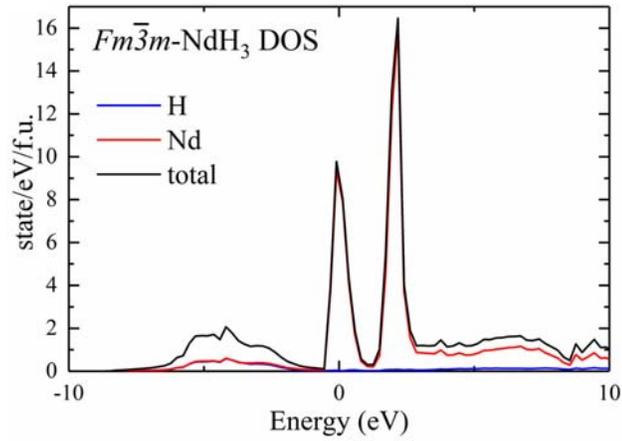

**Fig. S4.** Electronic density of states of the cubic $Fm\bar{3}m$-NdH$_3$ at 30 GPa.

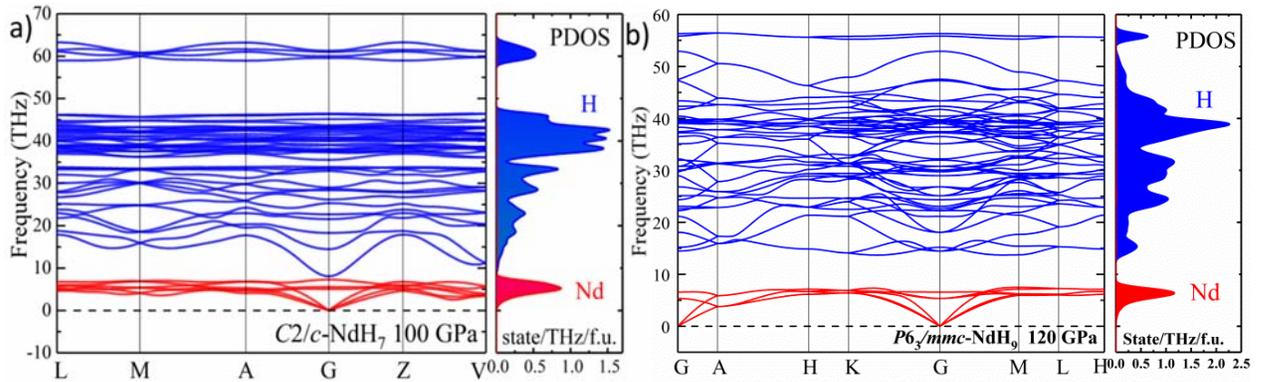

**Fig. S5.** Phonon band structure of **(a)** $C2/c$-NdH$_7$ at 100 GPa and **(b)** $P6_3/mmc$-NdH$_9$ at 120 GPa.

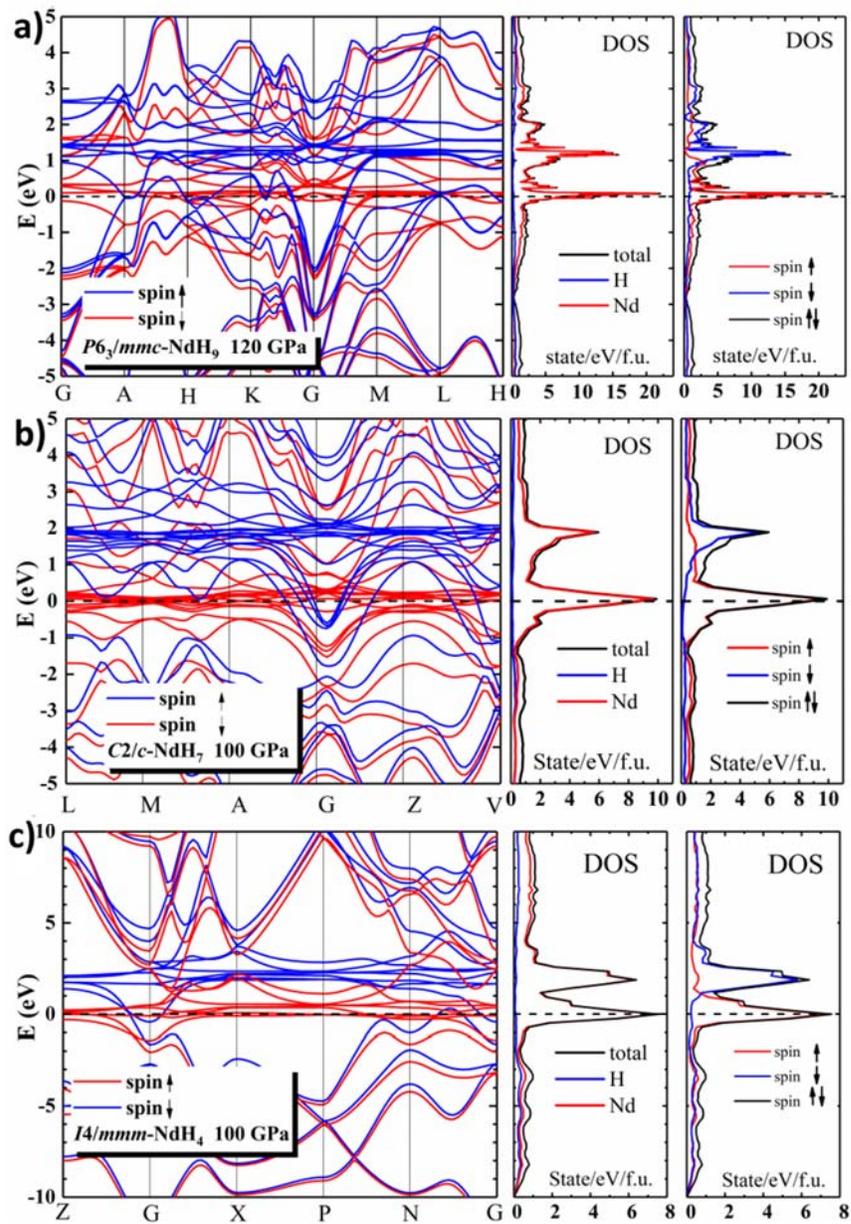

**Fig. S6.** Calculated band structure and densities of electron states in **(a)** $P6_3/mmc$-NdH$_9$, **(b)** $C2/c$-NdH$_7$ and **(c)** $I4/mmm$-NdH$_4$ with spin up (blue), spin down (red) and both spins (black) at 120 and 100 GPa.

# Magnetic properties of neodymium and praseodymium hydrides

In order to predict the magnetic ordering of *I4/mmm*-NdH$_4$ at 100 GPa and compare it with isostructural PrH$_4$[25], we have performed structure relaxation of 13 different magnetic configurations (Figure S7a) without spin-orbit coupling (SOC).

For NdH$_7$, structures of 10 different magnetic configurations (Figure S7b) were optimized at 100 GPa without SOC, while for NdH$_9$, structures of 6 different magnetic configurations (Figure S7c) were optimized at 120 GPa without SOC. Parameters of calculations were checked for convergence. The kinetic energy cutoff of the plane wave basis set was chosen to be 500 eV (520 eV – for NdH$_7$ and NdH$_9$) the length l = 40 (30 – for NdH$_7$ and NdH$_9$) for the automatic generation of Γ-centered Monkhorst-Pack grids as implemented in the VASP code, and the smearing parameter σ = 0.1 with the Methfessel-Paxton first-order method, which gave us a maximal error of 1 meV/atom with respect to more accurate calculations.

Our trial set of magnetic configurations involved one ferromagnetic (FM) and 12 anti-ferromagnetic (AFM) structures for Nd/PrH$_4$ (Figure S7a), 9 AFM configurations for NdH$_7$ (Figure S7b) and 5 AFM configurations for NdH$_9$ (Figure S7c). The AFM configurations are exhaustive up to supercell with four metal atoms and were generated using derivative structure enumeration library ENUMLIB[26]. In order to speed up calculations, the size of the unit cell has always been chosen to be the minimum size capable to represent the desired magnetic configuration: if we consider Nd or Pr atoms with different spin states to be different atomic types, we always worked with primitive cells. Relaxation of our trial set with fixed magnetic moments led to a loss of magnetic ordering for AFM 5, 6, 8 structures for PrH$_4$, AFM 6,8 – for NdH$_4$, and the final enthalpies of all configurations are given in Table S8.

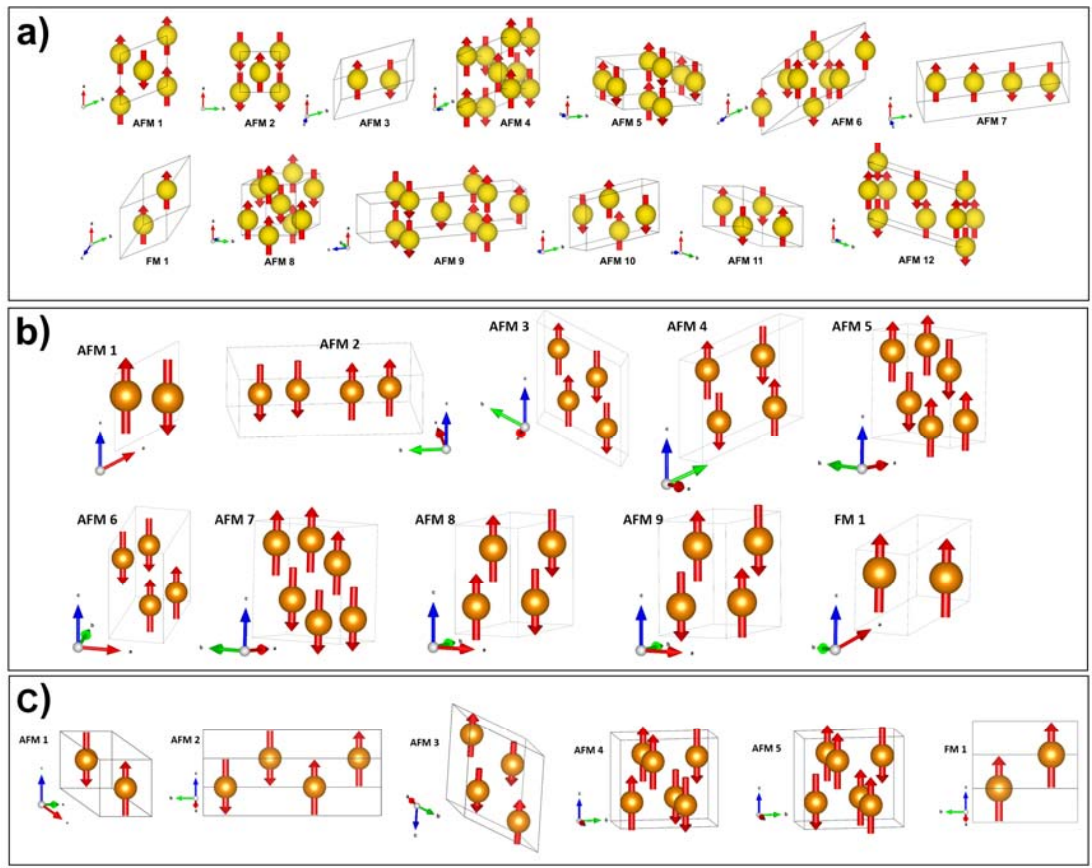

**Fig. S7.** Trial magnetic configurations for NdH$_4$/PrH$_4$ **(a)**, NdH$_7$ **(b)** and NdH$_9$ **(c)** used in this work. Note that different configurations can be represented in different unit cell settings as given by the ENUMLIB output.

**Table S8.** Relative enthalpies (meV per metal atom) of the magnetic configurations after relaxation without SOC for *I4/mmm*-PrH₄, *I4/mmm*-NdH₄, *C2/c*-NdH₇ at 100 GPa and *P6₃/mmc*-NdH₉ at 120 GPa.

| Configuration | FM 1 | AFM 1 | AFM 2 | AFM 3 | AFM 4 | AFM 5 | AFM 6 | AFM 7 | AFM 8 | AFM 9 | AFM 10 | AFM 11 | AFM 12 |
|---|---|---|---|---|---|---|---|---|---|---|---|---|---|
| Enthalpy PrH₄ | 29.80 | 89.25 | 54.45 | 71.40 | 142.25 | 279.25* | 283.70* | 86.35 | 287.15* | **0.0** | 137.40 | 60.55 | 99.90 |
| Enthalpy NdH₄ | 113.31 | 166.15 | **0.0** | 131.75 | 164.10 | 128.20 | 1151* | 90 | 1174.5* | 30.75 | 142.90 | 36.10 | 174.55 |
| Enthalpy NdH₇ | 95.85 | **0.0** | 37.74 | 100.65 | 94.17 | 113.60 | 53.85 | 52.13 | 89.15 | 84.13 | - | - | - |
| Enthalpy NdH₉ | 233.8 | **0.0** | 148.9 | 107.2 | 68.7 | 154.8 | - | - | - | - | - | - | - |

*these configurations turned out to be non-magnetic after relaxation.

After relaxation, the magnetic anisotropy of all configurations was investigated. Starting from the relaxed geometry of each configuration, we calculated the enthalpies of structures with magnetic moments aligned along different directions (X, Y, Z, XY, XZ, YZ, XYZ), taking into account SOC for each magnetic ordering. The results are summarized in Table S9. For 7 configurations the most favorable direction for magnetic moments is along the Z-axis for PrH₄ and the XYZ - for NdH₄, but the energy differences with the other alignments are too small to expect a macroscopic magnetic anisotropy.

Néel temperatures were estimated using mean-filed approximation $T_N^{MF} = \sum_{i,j} \frac{J_{i,j}S^2}{3k_B} \approx \min \frac{|E_{FM}-E_{AFM}|}{6k_B}$ where the energy difference at the numerator has been calculated taking into account SOC and using the relaxed geometry of the most stable magnetic configuration that we found, that is AFM 9, AFM 2, AFM 1 and AFM 1 respectively for PrH₄, NdH₄, NdH₇ and NdH₉, We got 20 K for PrH₄, 4 K for NdH₄, 251 K for NdH₇ and 136 K for NdH₉. It must be noted that, for PrH₄, SOC led to a higher stability of the FM configuration over the AFM 9 (the energy difference was calculated on the relaxed geometry of AFM 9).

**Table S9.** Enthalpies (meV per metal atom) of the magnetic configurations with the magnetic moments aligned along seven different directions. For each row, enthalpies are expressed with respect to the most stable alignment in that row. The number of atoms in the formula unit, including hydrogen, should be taken into account.

|  | Configuration | X | Y | Z | XY | XZ | YZ | XYZ |
|---|---|---|---|---|---|---|---|---|
| PrH₄ | FM | 0.125 | 0.125 | 0.050 | 0.050 | 0.040 | 0.040 | 0.000 |
|  | AFM 1 | 0.555 | 0.505 | 0.315 | 0.000 | 0.490 | 0.490 | 0.295 |
|  | AFM 2 | 1.875 | 1.875 | 0.000 | 1.735 | 0.620 | 0.620 | 0.955 |
|  | AFM 3 | 0.735 | 0.000 | 0.165 | 0.310 | 0.025 | 1.510 | 1.190 |
|  | AFM 4 | 0.780 | 0.780 | 0.000 | 0.320 | 0.280 | 0.245 | 0.135 |
|  | AFM 5 | 0.950 | 0.000 | 0.685 | 0.305 | 0.810 | 0.355 | 0.435 |
|  | AFM 6 | 1.395 | 1.415 | 0.000 | 1.495 | 0.390 | 0.495 | 0.735 |
|  | AFM 7 | 0.835 | 1.345 | 0.000 | 0.590 | 0.555 | 1.675 | 1.260 |
|  | AFM 8 | 1.575 | 1.570 | 0.000 | 1.520 | 0.315 | 0.505 | 0.665 |
|  | **AFM 9** | **1.920** | **1.920** | **0.000** | **1.490** | **0.865** | **0.940** | **1.270** |
|  | AFM 10 | 1.300 | 0.965 | 0.000 | 1.065 | 0.035 | 0.295 | 0.135 |
|  | AFM 11 | 1.990 | 0.605 | 0.135 | 1.065 | 0.000 | 0.200 | 0.180 |
|  | AFM 12 | 0.140 | 0.120 | 0.240 | 0.000 | 0.215 | 0.210 | 0.155 |
| NdH$_4$ | FM | 3.160 | 3.155 | 0.000 | 5.630 | 1.080 | 0.395 | 0.810 |
|  | AFM 1 | 10.990 | 10.770 | 0.015 | 11.700 | 0.000 | 0.195 | 0.540 |
|  | AFM 2 | **0.925** | **0.925** | **1.000** | **0.570** | **0.115** | **0.115** | **0.000** |
|  | AFM 3 | 0.000 | 21.425 | 8.245 | 0.115 | 0.195 | 7.955 | 0.460 |
|  | AFM 4 | 5.720 | 5.715 | 0.000 | 7.335 | 1.770 | 0.805 | 0.635 |
|  | AFM 5 | 140.595 | 0.000 | 14.095 | 0.095 | 0.865 | -* | - |
|  | AFM 6 | 1.505 | 1.250 | 0.000 | 2.175 | 0.710 | 0.535 | 1.695 |
|  | AFM 7 | 0.000 | 2.970 | 1.855 | - | - | - | - |
|  | AFM 8 | 6.140 | 6.095 | 0.000 | 8.605 | 3.260 | 0.760 | - |
|  | AFM 9 | 171.100 | 2.720 | 169.920 | 0.000 | 169.520 | 170.375 | - |
|  | AFM 10 | 16.865 | 4.190 | 0.125 | 9.085 | 6.445 | 0.000 | - |
|  | AFM 11 | 0.000 | 4.285 | 2.300 | 5.855 | 1.965 | 3.165 | - |
|  | AFM 12 | 141.955 | 141.985 | 131.210 | 140.485 | 131.505 | 0.000 | 135.200 |
| NdH$_7$ | FM | 4.981 | 7.7 | 0 | 3.729 | 5.513 | 0.083 | 1.697 |
|  | AFM 1 | 3.286 | 6.558 | 0 | 2.822 | 0.345 | 2.027 | - |
|  | AFM 2 | 2.778 | 7.074 | 0.232 | 2.735 | 0 | 2.038 | - |
|  | AFM 3 | 1.872 | 1.891 | 0.975 | 1.632 | 0.289 | 0.377 | 0 |
|  | AFM 4 | 1.113 | 5.397 | 1.097 | 2.823 | 0 | 3.399 | 1.056 |
|  | AFM 5 | 0.463 | 3.633 | 1.571 | 0 | 1.675 | 1.995 | 0.901 |
|  | AFM 6 | 2.599 | 10.267 | 0 | 5.798 | 1.441 | 0.984 | 1.203 |
|  | AFM 7 | 3.982 | 0 | 5.062 | 1.75 | 2.593 | 3.783 | - |
|  | AFM 8 | 5.944 | 0 | 5.253 | 3.355 | 0.516 | - | - |
|  | AFM 9 | 0 | 1.097 | 5.927 | 1.242 | 1.049 | - | - |
| NdH$_9$ | FM | 0.14 | 0.0 | 11.94 | 0.07 | 1.38 | - | 0.08 |
|  | AFM 1 | 0.01 | 0.01 | - | 0.0 | 1.56 | 2.41 | 0.95 |
|  | AFM 2 | 0.02 | 0.0 | 4.64 | 0.01 | - | 2.41 | 1.44 |
|  | AFM 3 | 1.7 | 0.4 | - | 1.03 | 3.01 | 0.0 | 0.70 |
|  | AFM 4 | 2.4 | 0.0 | 2.4 | 1.0 | 2.45 | 1.20 | 1.64 |
|  | AFM 5 | 0.0 | 8.84 | 12.04 | - | 2.38 | 10.08 | 3.16 |

*- missing values are calculations which do not converge for 200 steps

Recently Peña-Alvarez et al.[25] described high-pressure synthesis and equation of state of new tetragonal praseodymium hydride, PrH$_{4+x}$ with unit cell volume ~ 25 Å$^3$/Pr-atom which is significantly higher than the calculated volume of $I4/mmm$-PrH$_4$ in its most stable magnetic configuration: 23.07 Å$^3$ per Pr atom (100 GPa). The volume does not change either with or without SOC. This is about 2-2.5 Å$^3$/Pr-atom lower than the experimental values[25] and it is the largest volume among all magnetic configurations that were relaxed. The smallest volume is given by the AFM configurations 5, 6 and 8, which relaxation led to a non-magnetic state with minimum volume of 22.7 Å$^3$ per Pr atom at the same pressure.

The puzzle can be solved by taking into account strong correlations using DFT+U with Hubbard correction term $U$-$J$. This term has a negative impact on convergence and only 4 magnetic configurations from the trial set were successfully relaxed. The most stable configuration for PrH$_4$ at 100 GPa with $U$-$J$ = 3, 4 or 5 eV is AFM 2. Its volume with U = 3 eV at 100 GPa is 24.77 Å$^3$/Pr-atom, very close to the experimental one (Figure S8). Neel temperature of PrH$_4$ with U= 3 eV was estimated using mean-filed approximation as 37 K for AFM 2 configuration. For all 4 investigated magnetic configurations we found a maximum variation in the volume of 0.03 Å$^3$/Pr-atom. The alignment of magnetic moments along different directions has a negligible effect on the free energy of PrH$_4$, and therefore this material can be considered isotropic with respect to magnetic moments.

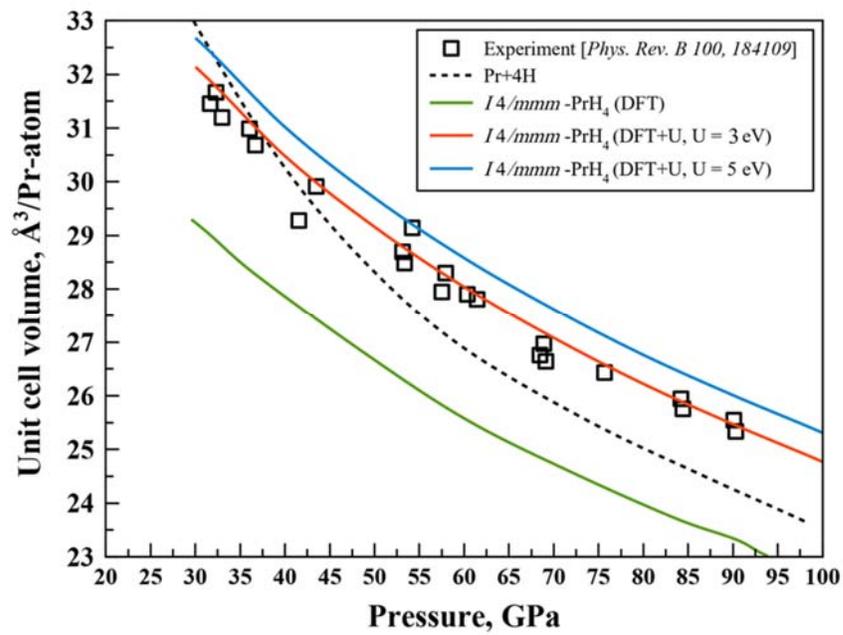

**Fig. S8.** Equation of state of *I4/mmm*-PrH₄ calculated using DFT and DFT+U with $U-J$ = 3 and 5 eV.

## Superconductivity in neodymium hydrides: Quantum ESPRESSO

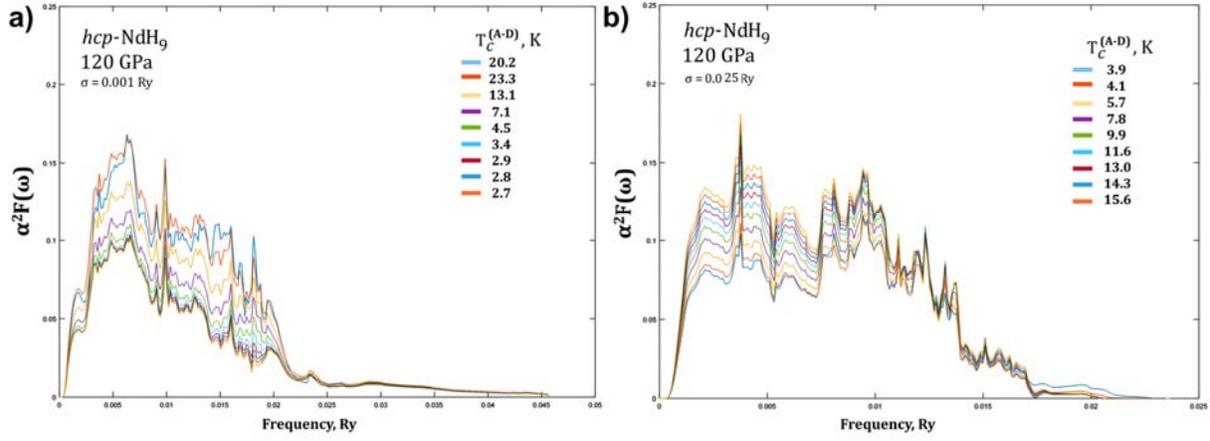

**Fig. S9.** A series of the Eliashberg functions calculated for *hcp*-NdH$_9$ (120 GPa) at different σ-broadening: **(a)** σ = 0.001…0.009 Ry, **(b)** σ = 0.025…0.225 Ry (μ*=0.1).

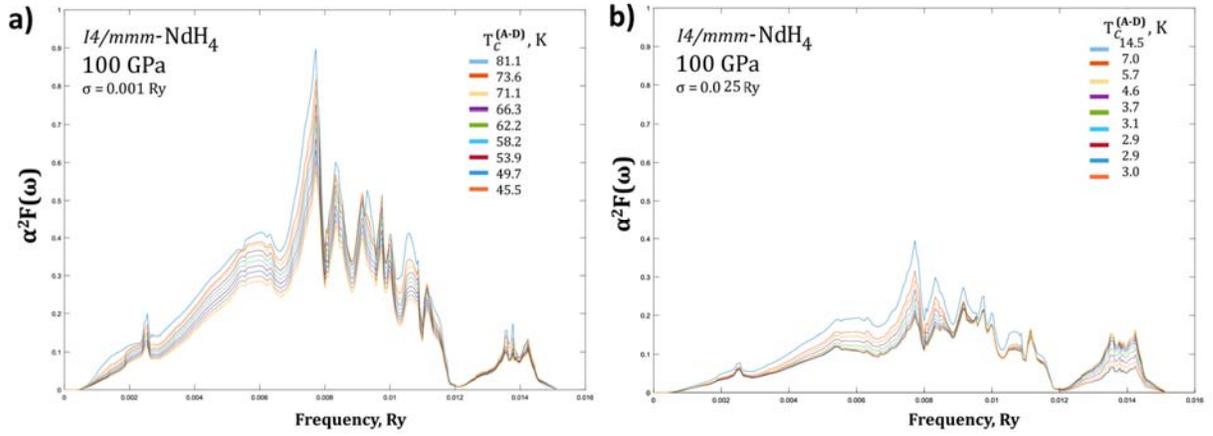

**Fig. S10.** A series of the Eliashberg functions calculated for *I4/mmm*-NdH$_4$ (100 GPa) at different σ-broadening: **(a)** σ = 0.001…0.009 Ry, **(b)** σ = 0.025…0.225 Ry. (μ*=0.1).

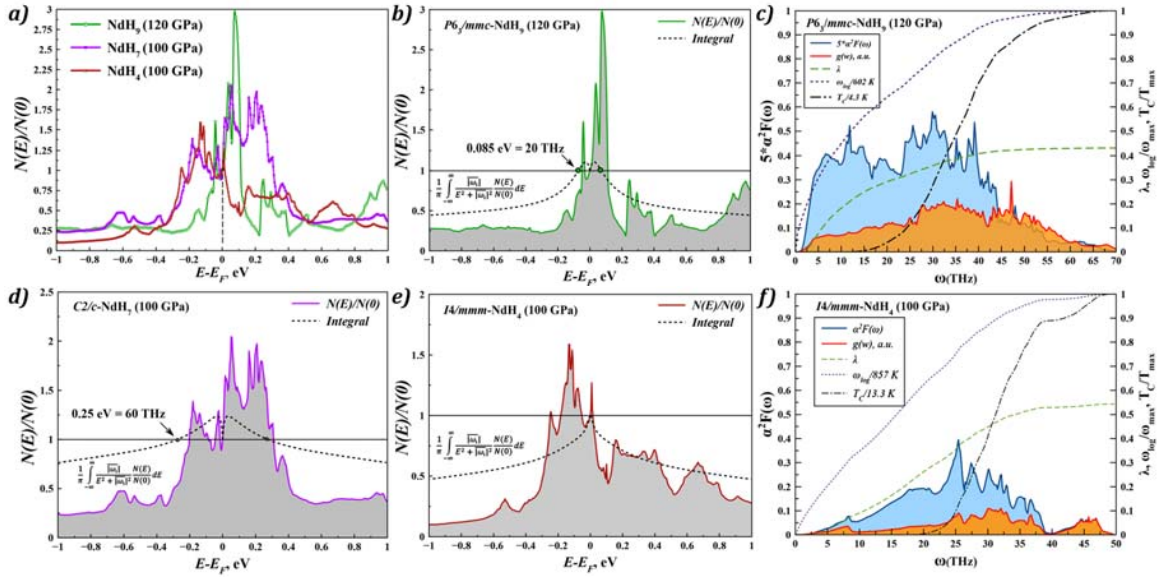

**Fig. S11.** *(a)* Comparative density of electronic states of the studied NdH$_x$ hydrides; *(b, d, e)* A series of $\overline{N}(\overline{|\omega_l|})$ integrals calculated for NdH$_4$, NdH$_7$ and NdH$_9$; *(c, f)* the Eliashberg functions at σ = 0.025 Ry calculated by Quantum ESPRESSO.

The critical temperature of superconducting transition with accounting variable density of electron states *N(E)* was obtained using the data calculated in Quantum ESPRESSO (Fig. S11) and Matsubara-type linearized Eliashberg equations, modified by Lie and Carbotte[27-28]

$$\hbar\omega_j = \pi(2j+1)\cdot k_B T, \quad j=0,\pm 1,\pm 2,... \quad (S2)$$

$$\lambda(\omega_i - \omega_j) = 2\int_0^\infty \frac{\omega \cdot \alpha^2 F(\omega)}{\omega^2 + (\omega_i - \omega_j)^2} d\omega \quad (S3)$$

$$\overline{N}(\overline{|\omega_l|}) = \frac{1}{\pi} \int_{-\infty}^\infty \frac{\overline{|\omega_l|}}{E^2 + \overline{|\omega_l|}^2} \frac{N(E)}{N(0)} dE \quad (S4)$$

$$\Delta(\omega=\omega_i,T) = \Delta_i(T) = \pi k_B T \sum_j \frac{\left[\lambda(\omega_i-\omega_j) - \mu^*\right]\cdot \overline{N}(|\overline{\omega_j}|)}{\rho + \left|\hbar\omega_j + \pi k_B T \sum_k (sign\,\omega_k)\cdot \lambda(\omega_j - \omega_k)\cdot \overline{N}(|\overline{\omega_k}|)\right|} \cdot \Delta_j(T) \quad (S5)$$

where *T* is temperature in Kelvins, $\mu^*$ is Coloumb pseudopotential, $\omega$ is frequency in Hz, $\rho(T)$ is a pair-breaking parameter, function $\lambda(\omega_i - \omega_j)$ relates to effective electron-electron interaction *via* exchange of phonons[29], *N(E)/N(0)* – dimensionless density of electron states. Transition temperature can be found as a solution of equation $\rho(T_C) = 0$ where $\rho(T)$ is defined as *max(ρ)* providing that *Δ(ω)* is not a zero function of *ω* at fixed temperature.

These equations can be rewritten in a matrix form as[30]

$$\rho(T)\psi_m = \sum_{n=0}^N K_{mn}\psi_n \Leftrightarrow \rho(T)\begin{pmatrix}\psi_1\\ \ldots\\ \psi_N\end{pmatrix} = \begin{pmatrix}K_{11} & \ldots & K_{1N}\\ \ldots & K_{ii} & \ldots\\ K_{N1} & \ldots & K_{NN}\end{pmatrix} \times \begin{pmatrix}\psi_1\\ \ldots\\ \psi_N\end{pmatrix}, \quad (S6)$$

where $\psi_n$ relates to *Δ(ω, T)*, and

$$K_{mn} = \{F(m-n) - \mu^*\}\cdot \overline{N}(|\overline{\omega_n}|) + \{F(m+n+1) - \mu^*\}\cdot \overline{N}(|\overline{\omega_{n+1}}|) - \delta_{mn}\left[2m+1+F(0)\cdot\overline{N}(|\overline{\omega_m}|) + 2\sum_{l=1}^m F(l)\cdot\overline{N}(|\overline{\omega_l}|)\right] \quad (S7)$$

$$F(x) = F(x,T) = 2\int_0^{\omega\max} \frac{\alpha^2 F(\omega)}{(\hbar\omega)^2 + (2\pi\cdot kT\cdot x)^2} \hbar^2 \omega d\omega, \quad (S8)$$

where $\delta_{nn} = 1$ and $\delta_{nm} = 0$ ($n \neq m$) – is a unit matrix. Now one can replace criterion of $\rho(T_C) = 0$ by the vanishing of the maximum eigenvalue of the matrix $K_{nm}$.

# Details of calculation of Eliashberg function for NdH$_9$: VASP & QE

It is possible to calculate the electron-phonon coupling in *hcp*-NdH$_9$ within Density Functional Perturbation Theory using the QE package. In doing so, we observed that phonon frequencies, $\omega_\nu(\mathbf{q})$, and the corresponding $\lambda_\nu(\mathbf{q})$ values turned out negative for many of the irreducible BZ q-points. This is also reflected in the fact that $\lambda(\mathbf{q})$ is not converged within the broad range of varied values of the Gaussian broadening parameter $\sigma$ (see Fig. S12). By convergence, we mean that the slope of any function $f(\sigma)$ goes to zero at some intermediate value within the varied $\sigma$ parameter range.

However, we observed that this deficiency does not appear in the values of the calculated phonon lifetime, $\gamma_\nu(\mathbf{q})$. For this quantity, QE calculations provide positive values and convergence is indeed achieved for $\sigma = 0.02$ Ry (Fig. S12b).

Phonon frequency calculations were also performed with VASP (Fig. S5(b)). In this case, all modes are positive and well behaving, indicating the stability of *hcp*-NdH$_9$ at 120 GPa. Using the equation, $\lambda_\nu(\mathbf{q}) = \frac{\gamma_\nu(\mathbf{q})}{\pi N_F \omega_\nu^2(\mathbf{q})}$ $\lambda_\nu(q) = \frac{\gamma_\nu(q)}{\pi N_F \omega_\nu^2(q)}$ where $\lambda_\nu(q) = 0$ for the first three (acoustic) modes, we combined the converged $\gamma_\nu(\mathbf{q})$ values from QE with the converged calculated $\omega_\nu(\mathbf{q})$ values from VASP and calculate a $\lambda_\nu(\mathbf{q})$ which is definite positive and yields an average value of $\lambda = 2.82$. In doing so, we firstly calculated $\omega_\nu(\mathbf{q})$ with VASP on a fine q-grid and subsequently interpolated the acquired values on the 4x4x2 q-grid where the electron-phonon calculations were performed in QE. With the obtained set of the combined $\lambda_\nu(\mathbf{q})$'s and the $\omega_\nu(\mathbf{q})$'s (calculated by VASP), we proceed to calculate the corresponding Eliashberg function using the formula: $\alpha^2 F(\omega) = \frac{1}{2}\sum_{q,\nu} \lambda_\nu(q) \cdot \omega_\nu(q) \cdot \delta(\omega - \omega_\nu(q))$ (Fig. S13). The respective logarithmic frequency can be calculated from the Eliashberg function, as well, and in this case was found as $\omega_{log}$ = 272.12 K. We used these data as input to our Eliashberg theory calculations for the superconducting state (see below).

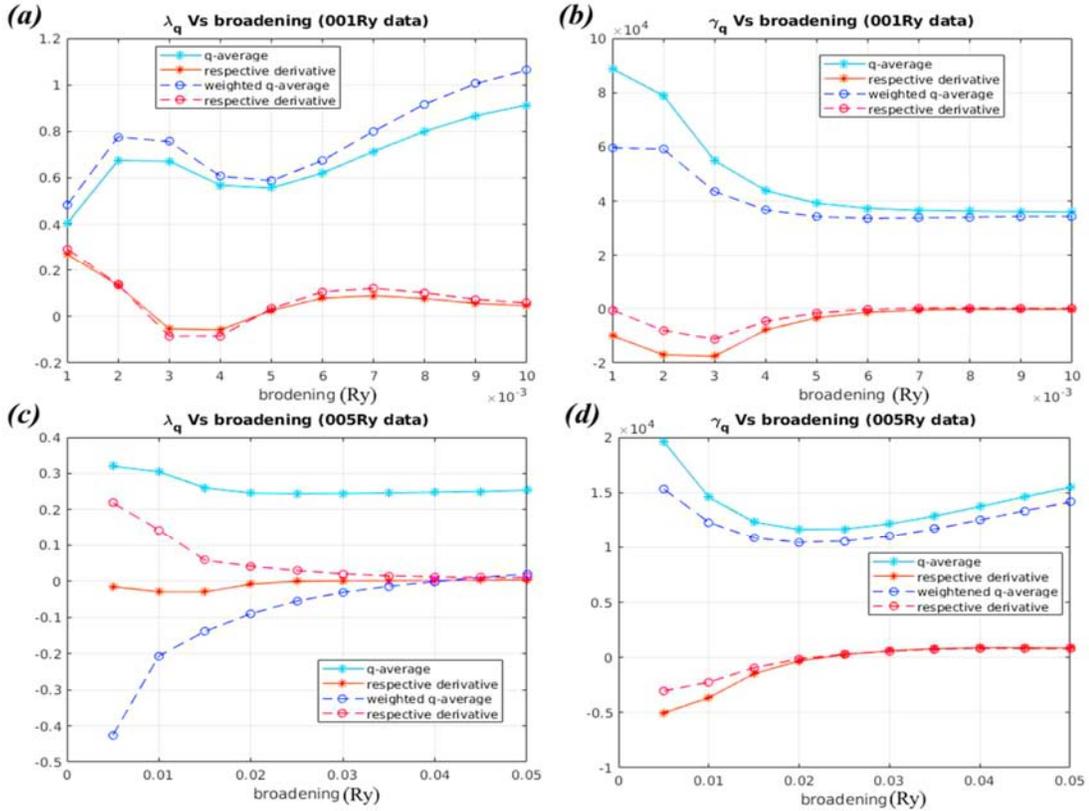

**Fig. S12.** Dependence of EPC coefficients ($\lambda_q$) and phonon lifetimes ($\gamma_q$, GHz) of NdH$_9$ at 120 GPa on the Gaussian broadening($\sigma$) calculated in QE using 2 steps of broadening: 0.001 Ry *(a, b)* and 0.005 Ry *(c, d)*. Convergence is achieved for $\lambda_q$ at $\sigma > 0.025$ Ry *(c)*, and for $\gamma_q$ at $\sigma > 0.01$ Ry *(b)*.

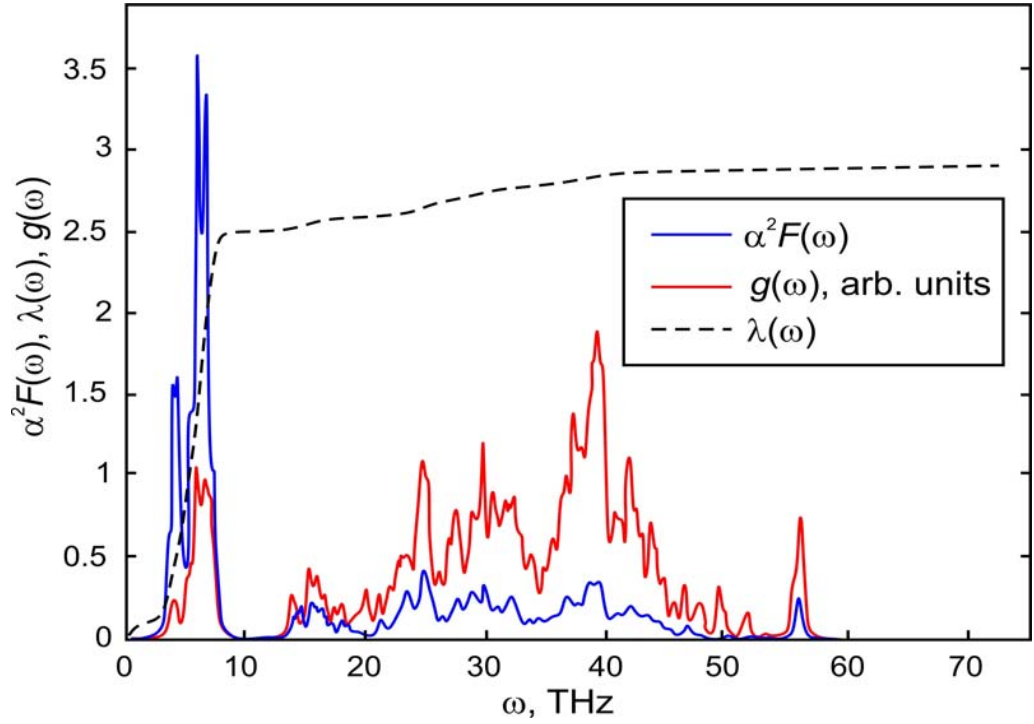

**Fig. S13.** Eliashberg function (blue), phonon density of states (red) and averaged $\lambda(\omega)$ of *hcp*-NdH$_9$ at 120 GPa reconstructed combining results of Quantum ESPRESSO and VASP. The spectral weight in $\alpha^2F$ is transferred from the high frequency H-modes to the low frequency Nd-modes due to high anisotropy of electron-phonon interaction in NdH$_9$ and the dominance of Nd-contribution to the electron DOS at $E_F$.

# Superconductivity in *P*6$_3$/*mmc*-NdH$_9$: UppSC code

For simplicity, assume a single band model of a collinear AFM state in the absence of SOC. In the unfolded Brillouin Zone the relevant Hamiltonian describing the electron kinetic part reads, $H = \sum_{k,\sigma} \xi_\sigma(\mathbf{k}) \cdot c^\dagger_{k\sigma} \cdot c_{k\sigma}$, where $\xi_\sigma(\mathbf{k})$ is the spin-resolved electron energy dispersion and σ = ±1. The spin-resolved band structure can be re-written as $\xi_\sigma(\mathbf{k}) = \xi(\mathbf{k}) - \sigma h(\mathbf{k})$ where we have introduced the Kramers degenerate electron energy dispersion, $\xi(\mathbf{k})$, and the effective spin-splitting, $h(\mathbf{k})$. The Hamiltonian can be written equivalently as: $H = \sum_{k,\sigma}[\xi(\mathbf{k}) - \sigma h(\mathbf{k})] \cdot c^\dagger_{k\sigma} \cdot c_{k\sigma}$, with $h(\mathbf{k}) = \frac{\xi_\uparrow(\mathbf{k}) - \xi_\downarrow(\mathbf{k})}{2}$ and $\xi(\mathbf{k}) = \frac{\xi_\uparrow(\mathbf{k}) + \xi_\downarrow(\mathbf{k})}{2}$. Writing the kinetic electron part of the Hamiltonian in this way, allows for the direct application of the recently developed theoretical framework for the *ab initio* modeling of superconductivity under magnetic fields [31] by solving the Eliashberg equations (S9)-(S12). Therefore, the effect of the magnetic state on the possible superconducting state is taken into account. Since, the Eliashberg theory employed here is formulated by integrating out electron degrees of freedom away from the Fermi surface, the relevant effective spin-splitting that enters in the theory is calculated for each band at the Fermi level (**k** = **k**$_F$) and subsequently we take the average, so that we end up with the parameter h ≈ 450 meV (or 890 meV for *U-J*) that we used in our self-consistent calculations. We note that the absolute sign of *h* does not play any role in the solution of Eq. (S9)-(S12) and that we have chosen the convention of *h* > 0.

To take into account possible effect of the magnetic structure on electron-phonon interaction we solved the following set of coupled self-consistent Eliashberg equations using as input the *ab initio* calculated electron, phonon and electron-phonon properties of *P*6$_3$/*mmc*-NdH$_9$ at 120 GPa (Figs. S5, S6 and S13).

$$Z(i\omega_n) = 1 + \frac{\pi T}{2\omega_n} \sum_{n',\pm} \lambda(\omega_n - \omega_{n'}) \frac{\omega_{n'} \pm i\tilde{H}(i\omega_{n'})}{\Theta(i\omega_{n'})}, \tag{S9}$$

$$\Sigma_h(i\omega_n) = \frac{\pi T}{2} \sum_{n',\pm} \lambda(\omega_n - \omega_{n'}) \frac{\tilde{H}(i\omega_{n'}) \mp i\omega_{n'}}{\Theta(i\omega_{n'})}, \tag{S10}$$

$$Z(i\omega_n)\Delta_e(i\omega_n) = \frac{\pi T}{2} \sum_{n',\pm} [\lambda(\omega_n - \omega_{n'}) - \mu^*(\omega_c)] \frac{\Delta_e(i\omega_{n'}) \mp i\Delta_o(i\omega_{n'})}{\Theta(i\omega_{n'})} \tag{S11}$$

$$Z(i\omega_n)\Delta_o(i\omega_n) = \frac{\pi T}{2} \sum_{n',\pm} [\lambda(\omega_n - \omega_{n'}) - \mu^*(\omega_c)] \frac{\Delta_o(i\omega_{n'}) \pm i\Delta_e(i\omega_{n'})}{\Theta(i\omega_{n'})} \tag{S12}$$

with

$$\Theta(i\omega_{n'}) = \left[\left(\omega_{n'} \pm i\tilde{H}(i\omega_{n'})\right)^2 + (-\Delta_e(i\omega_{n'}) \pm i\Delta_o(i\omega_{n'}))^2\right]^{\frac{1}{2}}$$

and $\tilde{H}(i\omega_n) = [\Sigma_h(i\omega_n) + \mu_B h]/Z(i\omega_n)$.

In the above, $Z(i\omega_n)$, is the mass renormalization function, $h$ and $\Sigma_h(i\omega_n)$ denote the effective magnetic field due to spin splitting and the corresponding self-energy term, respectively. $\Delta_e(i\omega_n)$ is the superconducting gap function for standard s-wave, spin singlet pairing with even frequency dependence. Lastly, $\Delta_o(i\omega_n)$ is the superconducting gap function with odd frequency, s-wave spin triplet symmetry. The latter term is included for completeness since it is known that under a finite magnetic field this will be induced, but is usually expected quite small [31].

The Matsubara dependent coupling kernel was calculated from the Eliashberg function by

$$\lambda(\omega_n - \omega_{n'}) = \int_0^\infty d\omega\, \alpha^2 F(\omega) \frac{2\omega}{(\omega_n - \omega_{n'})^2 + \omega^2}$$

As usual, for the Coulomb pseudopotential, $\mu^*(\omega_c)$, we assumed values in the range [0.1, 0.2] and a cut-off in the Matsubara frequency sum $\omega_c = 10 \times \omega_{log}$, so that convergence is achieved. We solved the above coupled Eliashberg equations iteratively allowing up to 3000-4000 iteration cycles and up to $\sim 9 \times 10^4$ Matsubara frequencies. A strict convergence criterion of a relative difference of $10^{-6}$ was imposed.

The effective magnetic field, $h$, was found from the spin resolved electronic band structure, $\xi_\sigma(\mathbf{k})$, as $h(\mathbf{k}_F) = \frac{\xi_\uparrow(\mathbf{k}_F) - \xi_\downarrow(\mathbf{k}_F)}{2}$. Near the Fermi level, this is almost momentum independent and has a value of ~0.45 eV (*U-J*=0), see e.g. Fig. S6(a). The results of the calculations are discussed in the main text.

# References

1. Eremets, M. I., Megabar high-pressure cells for Raman measurements. *Journal of Raman Spectroscopy* **2003,** *34* (7-8), 515-518.
2. Zhang, L. L.; Yan, S.; Jiang, S.; Yang, K.; Wang, H.; He, S.; Liang, D.; Zhang, L.; He, Y.; Lan, X.; Mao, C.; Wang, J.; Jiang, H.; Zheng, Y.; Dong, Z.; Zeng, L.; Li, A., Hard X-ray micro-focusing beamline at SSRF. *Nuclear Science And Techniques* **2015,** *26*, 060101.
3. Prescher, C.; Prakapenka, V. B., DIOPTAS: a program for reduction of two-dimensional X-ray diffraction data and data exploration. *High Pressure Research* **2015,** *35* (3), 223-230.
4. Young, R. A., The Rietveld Method. *International union of crystallography* **1993,** *5*, 1-38.
5. Petříček, V.; Dušek, M.; Palatinus, L., Crystallographic Computing System JANA2006: General features. *Z. Kristallogr.* **2014,** *229* (5), 345-352.
6. Bail, A. L.; Duroy, H.; Forquet, J. L., Ab-initio structure determination of LiSbWO6 by X-ray powder diffraction. *Mater. Res. Bull.* **1988,** *23*, 447-452.
7. Hohenberg, P.; Kohn, W., Inhomogeneous Electron Gas. *Phys. Rev.* **1964,** *136* (3B), B864-B871.
8. Kohn, W.; Sham, L. J., Self-Consistent Equations Including Exchange and Correlation Effects. *Phys. Rev.* **1965,** *140* (4A), A1133-A1138.
9. Perdew, J. P.; Burke, K.; Ernzerhof, M., Generalized Gradient Approximation Made Simple. *Phys. Rev. Lett.* **1996,** *77*, 3865-3868.
10. Blöchl, P. E., Projector augmented-wave method. *Phys. Rev. B* **1994,** *50* (24), 17953-17979.
11. Kresse, G.; Joubert, D., From ultrasoft pseudopotentials to the projector augmented-wave method. *Phys. Rev. B* **1999,** *59*, 1758.
12. Kresse, G.; Hafner, J., Ab initiomolecular dynamics for open-shell transition metals. *Phys. Rev. B* **1993,** *48* (17), 13115-13118.
13. Kresse, G.; Hafner, J., Ab initiomolecular-dynamics simulation of the liquid-metal–amorphous-semiconductor transition in germanium. *Phys. Rev. B* **1994,** *49* (20), 14251-14269.
14. Kresse, G.; Hafner, J., Efficient iterative schemes for ab initio total-energy calculations using a plane-wave basis set. *Phys. Rev. B* **1996,** *54*, 11169.
15. Hobbs, D.; Kresse, G.; Hafner, J., Fully unconstrained noncollinear magnetism within the projector augmented-wave method. *Phys. Rev. B* **2000,** *62*, 11558.
16. Francis, B., Finite Elastic Strain of Cubic Crystals. *Phys. Rev.* **1947,** *71* (11), 809-824.
17. Gonzalez-Platas, J.; Alvaro, M.; Nestola, F.; Angel, R., EosFit7-GUI: a new graphical user interface for equation of state calculations, analyses and teaching. *J. Applied Crystallography* **2016,** *49* (4), 1377-1382.
18. Togo, A.; Oba, F.; Tanaka, I., First-principles calculations of the ferroelastic transition between rutile-type andCaCl2-typeSiO2at high pressures. *Phys. Rev. B* **2008,** *78* (13), 134106.
19. Togo, A.; Tanaka, I., First principles phonon calculations in materials science. *Scripta Materialia* **2015,** *108*, 1-5.
20. Giannozzi, P.; Baroni, S.; Bonini, N.; Calandra, M.; Car, R.; Cavazzoni, C.; Ceresoli, D.; Chiarotti, G. L.; Cococcioni, M.; Dabo, I.; Corso, A. D.; Gironcoli, S. d.; Fabris, S.; Fratesi, G.; Gebauer, R.; Gerstmann, U.; Gougoussis, C.; Kokalj, A.; Lazzeri, M.; Martin-Samos, L.; Marzari, N.; Mauri, F.; Mazzarello, R.; Paolini, S.; Pasquarello, A.; Paulatto, L.; Sbraccia, C.; Scandolo, S.; Sclauzero, G.; Seitsonen, A. P.; Smogunov, A.; Umari, P.; Wentzcovitch, R. M., QUANTUM ESPRESSO: a modular and open-source software project for quantum simulations of materials. *J. Phys.: Condens. Matter* **2009,** *21* (39), 395502.
21. Baroni, S.; Gironcoli, S. d.; Corso, A. D.; Giannozzi, P., Phonons and related crystal properties from density-functional perturbation theory. *Rev. Mod. Phys* **2001,** *73*, 515-562
22. Allen, P. B.; Dynes, R. C., Transition temperature of strong-coupled superconductors reanalyzed. *Phys. Rev. B* **1975,** *12* (3), 905-922.
23. Chesnut, G. N.; Vohra, Y. K., a-uranium phase in compressed neodymium metal *Phys. Rev. B.* **2000,** *61*, R3768-R3771.
24. Pépin, C. M.; Geneste, G.; Dewaele, A.; Mezouar, M.; Loubeyre, P., Synthesis of FeH5: A layered structure with atomic hydrogen slabs. *Science* **2017,** *357*, 382.
25. Peña-Alvarez, M.; Binns, J.; Hermann, A.; Kelsall, L. C.; Dalladay-Simpson, P.; Gregoryanz, E.; Howie, R. T., Praseodymium polyhydrides synthesized at high temperatures and pressures. *Phys. Rev. B* **2019,** *100*, 184109.

26. Hart, G. L. W.; Nelson, L. J.; Forcade, R. W., Generating derivative structures at a fixed concentration. *Comp. Mat. Sci.* **2012,** *59*, 101-107.
27. Lie, S. G.; Carbotte, J. P., Dependence of Tc on electronic density of states. *Sol. State. Comm.* **1978,** *26* (8), 511-514.
28. Eliashberg, G. M., Interactions between Electrons and Lattice Vibrations in a Superconductor. *JETP* **1959,** *11* (3), 696-702.
29. Bergmann, G.; Rainer, D., The sensitivity of the transition temperature to changes in $\alpha^2 F(\omega)$ | SpringerLink. *Z. Für Phys.* **1973,** *263* (1), 59-68.
30. Allen, P. B.; Dynes, R. C. *A computer program for numerical solution of the Eliashberg equation to find Tc*; 1974; p TCM/4/1974
31. Aperis, A.; Maldonado, P.; Oppeneer, P. M., Ab initio theory of magnetic-field-induced odd-frequency two-band superconductivity in $MgB_2$. *Phys. Rev. B* **2015,** *92*, 054516.